\documentclass[aps,prd,showpacs,eqsecnum,twocolumn,superscriptaddress]{revtex4}
\usepackage{amsmath,amssymb}
\usepackage{graphicx}
\usepackage{color}

\begin{document}

\title{Three dimensional evolution of differentially rotating
magnetized neutron stars
}

\author{Kenta Kiuchi}
\affiliation{Yukawa Institute for Theoretical Physics,
Kyoto University, Kyoto, 606-8502, Japan~}

\author{Koutarou Kyutoku}
\affiliation{Yukawa Institute for Theoretical Physics,
Kyoto University, Kyoto, 606-8502, Japan~}
\affiliation{Theory Center, Institute of Particles and Nuclear Studies, KEK,
Tsukuba, Ibaraki, 305-0801, Japan~}

\author{Masaru Shibata}
\affiliation{Yukawa Institute for Theoretical Physics,
Kyoto University, Kyoto, 606-8502, Japan~}

\date{\today}

\begin{abstract}
We construct a new three-dimensional general relativistic
magnetohydrodynamics code, in which a fixed mesh refinement technique
is implemented. To ensure the divergence-free condition as well as the
magnetic flux conservation, we employ the method by
Balsara~\cite{Balsara:2001rw}. Using this new code, we evolve
differentially rotating magnetized neutron stars, and find that a
magnetically driven outflow is launched from the star exhibiting a
kink instability.  The matter ejection rate and Poynting flux are
still consistent with our previous finding~\cite{Shibata:2011fj}
obtained in axisymmetric simulations.
\end{abstract}

\pacs{04.25.D-, 04.30.-w, 04.40.Dg}

\maketitle

\section{Introduction}

Ground-based gravitational-wave detectors, Advanced LIGO,
Advanced VIRGO, and KAGRA will be in operation in the next five
years~\cite{Detectors}.
The first observation of gravitational waves, thus, will be achieved
in the near future. Among their sources of gravitational waves,
coalescence of binary neutron stars (BNS) is the most promising one,
and the detection of gravitational waves from them will provide us
unique information of strongly gravitational fields and properties of
dense nuclear matter. The BNS merger is also the potential candidate
for the progenitor of short-hard gamma-ray
bursts~\cite{Nakar:2007yr}. For the theoretical studies of the BNS
mergers, numerical relativity is the unique and robust approach.  A
number of numerical simulations have been
performed~\cite{Shibata:2003ga,Shibata:2005ss,Shibata:2006nm,Yamamoto:2008js,Kiuchi:2009jt,Kiuchi:2010ze,Hotokezaka:2011dh,Sekiguchi:2011mc,Sekiguchi:2011zd,Baiotti:2008ra,Rezzolla:2010fd,Baiotti:2011am,Marronetti:2003hx,Faber:2006qc,Duez:2009yz,Anderson:2007kz,Jana}
since the first success in 2000~\cite{Shibata:1999wm}.

Magnetic fields could play an important role in BNS mergers because
the inferred value of the magnetic-field strength via the observed
spin period $P$ and their time derivative $\dot{P}$ is high as
$10^{11}-10^{14}$ G for radio pulsars, of which more than 1800 are
known to date~\cite{Manchester:2004bp}.  During the merger process,
several mechanisms such as compression, magnetic winding, and
magnetorotational (MRI) instability~\cite{Balbus:1998ja} could amplify
their magnetic-field strength. This amplified magnetic field could
have an impact on the dynamics of the mergers because it contributes
to the angular momentum transport and the magnetic pressure may modify
the structure of the objects formed after the merger.  Motivated by
this expectation, several groups have implemented the
magnetohydrodynamics (MHD) code in the framework of numerical
relativity~\cite{Duez:2005sf,Shibata:2005gp,Giacomazzo:2007ti,Liebling:2010bn,Bucciantini:2010ax,Kuroda:2010yc}.
These numerical codes developed have been applied to collapse of
magnetized hypermassive neutron stars
(HMNS)~\cite{Duez:2005cj,Shibata:2005mz,Duez:2006qe}, magnetized
neutron star-black hole binary
merger~\cite{Chawla:2010sw,Etienne:2011ea}, evolution of magnetized
neutron stars~\cite{Kiuchi:2008ss,Kiuchi:2011yt,Shibata:2011fj}, and
magnetorotational collapse of massive stellar
cores~\cite{Shibata:2006hr,Kuroda:2010yc}.

In the context of BNS mergers, several groups have assessed what the
role of magnetic fields is during the inspiral and merger
~\cite{Anderson:2008zp,Liu:2008xy,Giacomazzo:2009mp,Giacomazzo:2010bx,Rezzolla:2011da}.
Their findings are summarized as follows: As long as the magnetic-field
strength before the merger is not unrealistically large, e.g.,
$10^{16}$--$10^{17}{\rm G}$, the magnetic field does not give a strong
impact on the inspiral dynamics.  When the external layers of the two
neutron stars come into contact, the Kelvin-Helmholtz instability
develops and forms vortexes.  Poloidal magnetic-field lines are curled
by them and generate a toroidal field in a short timescale.  The
saturation point of the magnetic-field strength is still under debate
because the field strength found in Ref.~\cite{Giacomazzo:2010bx} is
not as high as found in Ref.~\cite{Price:2006fi}. If the total
mass of BNS is large enough to collapse to a black hole surrounded by
a torus, the magnetic field in the torus may be subject to the
MRI~\cite{Rezzolla:2011da}.  On the other hand, if the total mass is
not large enough, a long-lived HMNS~\footnote{Long-lived HMNS implies
that its lifetime is much longer than the dynamical timescale $\sim
1$~s. For the detailed definition of the long-lived HMNS, see
Ref.~\cite{Hotokezaka:2011dh}} is born and the magnetic-field
amplification would be realized inside the HMNS. The later case has
not been explored in detail, because high computational costs for a
longterm well-resolved simulation prevent this study.

The recent measurement of mass for PSR J1614-2230
$(M_{J1916-2230}=1.97 \pm 0.04 M_\odot)$~\cite{Demorest:2010bx} gave a
strong constraint on nuclear equations of state (EOS). Together with
the fact that the canonical observed mass of neutron stars is
$1.3$--$1.4M_\odot$, it is natural to infer that a long-lived HMNS
will be born in the merger of BNS composed of neutron stars with the
canonical mass~\cite{Hotokezaka:2011dh}.  This implies that it is
mandatory to perform a longterm and high-resolution simulation of
magnetized BNS mergers.

In Ref.~\cite{Kiuchi:2011yt}, we have developed a three-dimensional
general relativistic magnetohydrodynamics (GRMHD) code, which has an
uni-grid structure with fish-eye coordinates. The dynamical range of
the BNS system is quite large spanning from the neutron-star size to
the wave length of gravitational waves.  Thus, we should implement a
mesh refinement technique to save computational costs. In the
framework of GRMHD, the implementation of mesh refinement techniques
has been done in three methods. For all the approaches, a special care
for preserving the divergence-free condition of magnetic fields is
taken.  In the first approach, equations for vector potentials instead
of magnetic fields are solved. In this method, any unconstrained
interpolations in the refinement boundaries, where the boundary
condition for child domains are determined using the data of their
parent domains, may be allowed for preserving the divergence-free
condition of magnetic
fields~\cite{Giacomazzo:2010bx,Etienne:2010ui,Etienne:2011re}. In the
second approach, the hyperbolic divergence-cleaning prescription is
employed to ensure the divergence-free condition of magnetic
fields~\cite{Anderson:2008zp,Liebling:2010bn}. The third
one~\cite{Kuroda:2010yc} is based on the Balsara's
constrained-transport scheme, in which a special interpolation scheme
in the refinement boundaries is mandatory to preserve the
divergence-free condition and {\em the magnetic-flux
conservation}~\cite{Balsara:2001rw,Balsara:2009}.  We construct a new
GRMHD code employing the third approach (modifying the original scheme
for the use in the vertex-centered grid) to precisely guarantee the
divergence-free condition and the magnetic-flux conservation. It is
worthy to note that this method is likely to work well also in the
presence of a black hole.

As the first application of our new code, we extend our previous work
in Ref.~\cite{Shibata:2011fj}, in which an axisymmetric HMNS with
magnetic fields was evolved. In that work, we found that a mildly
relativistic outflow is driven from the HMNS accompanying a strong
Poynting flux of magnitude proportional to $B^2 R^3 \Omega$ (where
$B$, $R$, and $\Omega$ denote the typical magnitudes of the magnetic
field, radius, and angular velocity of the HMNS) emitted toward the
direction of the rotational direction.  However, it was not clear that
three-dimensional effects, in particular the effect of nonaxisymmetric
instabilities such as kink instability~\cite{Goedbloed}, would not
play a role in this phenomenon. For a more physical study, we
obviously had to perform a three-dimensional simulation.

The paper is organized as follows: In Sec.~\ref{sec:formulation}, the
formulation to solve Einstein's equations as well as GRMHD equations
are briefly summarized. In addition, we briefly describe a method to
implement the fixed mesh refinement (FMR) algorithm in particular for
magnetic fields, and also mention the initial condition and grid
setup.  In Sec.~\ref{sec:result}, we present numerical results for the
evolution of a rapidly rotating magnetized neutron star, focusing on
the properties of the material and Poynting flux ejected from it.
Section~\ref{sec:summary} is devoted to discussing the implication of
our numerical results and a summary of this paper. In the Appendix, our
method for implementing the FMR scheme and results for the several
standard test-bed simulations are shown.  Throughout this paper, Greek
and Latin indices denote the spacetime and spatial components,
respectively.

\section{Formulation, method and model}\label{sec:formulation}
\subsection{Formulation and numerical issue}\label{sec:form}

We study the evolution of a rapidly rotating magnetized neutron star
by a three-dimensional GRMHD simulation in the framework of ideal MHD.
The formulation and numerical scheme for solving Einstein's equations
are the same as in Ref.~\cite{Kiuchi:2011yt}, in which one of the
Baumgarte-Shapiro-Shibata-Nakamura (BSSN)
formulations~\cite{Shibata:1995we,Baumgarte:1998te,Campanelli:2005dd,Baker:2005vv}
is employed, and a fourth-order finite-differencing scheme in the
spatial direction and a fourth-order Runge-Kutta scheme in the time
integration are implemented. The advection terms in Einstein's
evolution equations are evaluated with a fourth-order lopsided
finite-differencing scheme, as, e.g., in
Ref.~\cite{Brugmann:2008zz}. A conservative shock-capturing scheme is
employed to integrate GRMHD equations.  Specifically, we use a high
resolution central scheme~\cite{Kruganov:2000} with the third-order
piece-wise parabolic interpolation and a steep min-mod limiter.

We implement a FMR algorithm to our original three-dimensional GRMHD
code~\cite{Kiuchi:2011yt} which has an uni-grid structure with
fish-eye coordinates~\cite{Alcubierre:2002kk}. Our FMR scheme is
essentially the same as an adaptive-mesh refinement (AMR) scheme of
{\tt SACRA}~\cite{Yamamoto:2008js}, and enables us to assign fine grids
in the vicinities of neutron stars or black holes, while enlarging the
computational domain which covers a local wave zone of gravitational
waves with less computational cost.  The schemes for solving
Einstein's equations and hydrodynamics equations (the continuity,
momentum, and energy equations) are also the same as those of {\tt
SACRA} code~\cite{Yamamoto:2008js}, in which geometric variables and
fluid variables (density, pressure, internal energy, specific
momentum, and specific energy) are placed at the vertex-centered grids
and the grid spacing of a ``parent'' domain is twice as large as that
of its ``child'' domain. Each domain is equally composed of
$(2N+1,2N+1,2N+1)$ Cartesian grid zones for $(x,y,z)$, each of which
covers the interval $[-N\Delta x_{l},N\Delta x_{l}]$ for the $x$-,
$y$- and $z$-directions with $\Delta x_l$ being the grid spacing of
the $l$-th FMR level.  The label $l$ varies from 1 (for the coarsest
and largest domain) to $l_{\rm max}$ (for the finest and smallest
one).  The prolongation, i.e., interpolation from a ``child'' domain to a ``parent'' domain, 
of the geometric and fluid variables (i.e.,
the interpolation of the data of the parent domain for determining the
data in the child domains) in the refinement boundaries are done with
a fifth-order Lagrange interpolation. Because our grid is located at
the vertex centers, the restriction procedure, i.e., interpolation from a ``parent'' domain to a ``child'' domain, 
is straightforwardly done, by simply copying the data from a child domain to its parent
domain.

On the other hand, for integrating the induction equation, we need a
special care to preserve the divergence-free condition of magnetic
fields and to guarantee the magnetic-flux conservation. For
this purpose, several GRMHD codes constructed so
far~\cite{Giacomazzo:2007ti,Duez:2005sf,Shibata:2005gp,Etienne:2010ui}
have implemented either the constrained-transport
(CT)~\cite{Evans:1988qd} or flux-CT scheme~\cite{Balsara:1999}. In the
code of implementing FMR or AMR algorithm, we are required to take an
additionally special care when performing the prolongation and
restriction procedures of magnetic fields in the refinement
boundaries, as described in
Refs.~\cite{Balsara:2001rw,Balsara:2009,Etienne:2010ui,Etienne:2011re}.
AMR-GRMHD codes of Refs.~\cite{Etienne:2010ui,Giacomazzo:2010bx}
exploited a method of evolving the vector potential.  This method
guarantees the preservation of the divergence-free condition for
magnetic fields avoiding complex interpolation procedures in the
refinement boundaries.  However, {\em this method does not guarantee
the magnetic-flux conservation} in the refinement boundaries,
because the magnetic fields are calculated by a finite differencing of
the vector potential and this procedure does not in general guarantee
the magnetic-flux conservation.

Alternatively, the AMR-GRMHD code of Ref.~\cite{Liebling:2010bn}
employs a hyperbolic divergence-cleaning
technique~\cite{Anderson:2006ay}. In this scheme, a scalar field is
introduced, which is coupled to the system of the MHD and induction
equations. No special prescription is needed for the finite
differencing when solving GRMHD equations, and a non-zero divergence
of magnetic fields either propagates or damps away when they are
spuriously excited. However, as mentioned in
Ref.~\cite{Etienne:2010ui}, this method is likely to be incompatible
with the moving puncture method~\cite{Campanelli:2005dd,Baker:2005vv},
which is commonly used to evolve black-hole spacetimes in the BSSN
formulations.

The AMR-GRMHD code in Ref.~\cite{Kuroda:2010yc} implements the flux-CT
scheme for magnetic fields. In this scheme, both the preservation of
the divergence-free condition and the magnetic-flux conservation are
guaranteed in the refinement boundaries in the machine precision
level. However, the code of Ref.~\cite{Kuroda:2010yc} is second-order
accurate both in space and in time. We have developed a new code,
which is based on the flux-CT scheme (i.e., which can ensure the
magnetic-flux conservation and the divergence-free condition in the
refinement boundaries), and which is fourth-order accurate in time.
Following the method described in
Refs.~\cite{Balsara:2001rw,Balsara:2009}, we employ a
divergence-free-preserving interpolation based on
WENO5~\cite{Jiang:1996}.  In Appendix A, the grid structure, the
scheme for integrating the induction equation, as well as the
prolongation/restriction schemes of magnetic fields are described in
details. Results for several test-bed simulations in Appendix B
illustrate the reliability of our new code.

\subsection{Initial condition, density atmosphere and grid setup}


Following Refs.~\cite{Shibata:2005mz,Shibata:2011fj}, we adopt a
rapidly and differentially rotating neutron star in an axisymmetric
equilibrium as the initial conditions. It is a model of the HMNS
formed after the merger of a BNS. To model the neutron star, the
following piecewise polytropic EOS composed of two
pieces is employed:
\begin{eqnarray}
P_{\rm cold} = \Big\{
\begin{array}{ll}
K_1 \rho^{\Gamma_1} & (\rho \le \rho_{\rm nuc}),\\
K_2 \rho^{\Gamma_2} & (\rho \ge \rho_{\rm nuc}).\\
\end{array}
\end{eqnarray}
Here, $P$ and $\rho$ are the pressure and rest-mass density,
respectively. The specific internal energy, $\varepsilon$, is derived
assuming the first law of thermodynamics $d\varepsilon=-Pd(1/\rho)$,
and this specific internal energy written as a function of $\rho$ is
referred to as $\varepsilon_{\rm cold}$ (i.e., we initially set
$\varepsilon=\varepsilon_{\rm cold}(\rho)$).  Following
Ref.~\cite{Shibata:2005mz}, the parameters are chosen to be
$\Gamma_1=1.3$, $\Gamma_2=2.75$, $K_1=5.16\times 10^{14}$ cgs,
$K_2=K_1\rho_{\rm nuc}^{\Gamma_1-\Gamma_2}$, and $\rho_{\rm
nuc}=1.8\times 10^{14}~{\rm g/cm^3}$. This EOS produces spherical
neutron stars whose maximum gravitational mass $M_{\rm max}$ (rest
mass $M_{\rm b,max}$) is $2.01 M_\odot$ $(2.32M_\odot)$ and rigidly
rotating neutron stars with $M_{\rm max}~(M_{\rm b,max})=2.27
M_\odot~(2.60 M_\odot)$.  For the rotational law, we assume the same
profile as employed in
Refs.~\cite{Shibata:2005mz,Shibata:2011fj}. Table~\ref{tab:model}
shows the parameters of the differentially rotating neutron-star model
which we adopt.

During the simulations, we use a hybrid EOS as $P=P_{\rm
cold}+(\Gamma_{\rm th}-1)\rho(\varepsilon-\varepsilon_{\rm cold})$
with $\Gamma_{\rm th}=\Gamma_1$.  Our choice of $\Gamma_{\rm th}$ may
be rather small. We choose this small value to focus on the mass
ejection from the rotating neutron star primarily by the
magnetorotational effects suppressing shock heating effects.

A dipole magnetic field is superimposed initially. We assume that the
axis of the dipole is aligned with the rotation axis as in the
previous paper~\cite{Shibata:2011fj}, and write the vector potential in
the form
\begin{equation}
A_{\varphi}={A_0\varpi_0^2 \over (R^2+z^2+\varpi_0^2)^{3/2}},
\end{equation}
where we used the cylindrical coordinate $(R, z, \varphi)$. $\varpi_0$
is set to be $10/3 R_{\rm e}$ with $R_{\rm e}$ being the equatorial
stellar radius.  $A_0$ determines the magnetic-field strength and we
adjust this parameter to achieve the maximum field strength $B_0$ to
be $4.2\times 10^{13}~{\rm G}$ and $1.7 \times 10^{14}~{\rm G}$.
According to the magnetic-field strength, we refer to these models as
B13 and B14, respectively.  Here, $B_0$ is defined by $B \equiv
\sqrt{b^\mu b_\mu}$ where $b^\mu$ is the four-vector of the magnetic
field in the fluid rest frame.


We note that there is no reason to believe that the dipole axis is
aligned with the rotation axis for the HMNS formed after a BNS
merger. The reason for our choice of this simple profile is that the
purpose of this paper is to compare the results in three-dimensional
simulations with those in the axisymmetric one performed in
Ref.~\cite{Shibata:2011fj}. If the axes of the dipole and rotation do
not align with each other, the mechanism for the amplification of the
magnetic field and subsequent dynamical process of the system could be
significantly modified.  We will perform more systematic studies
varying the axis direction of the dipole in the future work.


As discussed in Ref.~\cite{Shibata:2011fj}, a tenuous density
atmosphere has to be put outside the neutron star for stably evolving
magnetically driven outflows.  If the atmosphere is dense, the outflow
and magnetic-field profile are substantially affected by the inertia 
of the atmosphere. Thus, we have to set the density of the atmosphere
to be as low as possible.  Specifically, we set it as
\begin{eqnarray}
 \rho_{\rm at}=\left\{
 \begin{array}{ll}
 f_{\rm at}~\rho_{\rm max}                   &~~(r\le 2 R_{\rm e}),\\
 f_{\rm at}~\rho_{\rm max}(r/2R_{\rm e})^{-n}&~~(r\ge 2 R_{\rm e}),\\
 \end{array}
 \right.
\end{eqnarray}
where $\rho_{\rm max}$ denotes the maximum rest-mass density of the
neutron star.  We set $f_{\rm at}=10^{-9}$ and $n=2$. As long as we
use these values, the magnetic field evolution depends only weakly on
the atmosphere~\cite{Shibata:2011fj}.

Table~\ref{tab:grid} summarizes the dipole field strength and grid
setup. The stars are contained in the numerical domain composed of the
finest grid resolution. In the typical simulations, $R_{\rm e}$ is
covered by 80 grid points in the finest domain. The side length in one
positive direction of the finest domain is $1.2 R_{\rm e}$.  We
prepare 8 refinement domains and in this case, the outer boundary is
located at $\approx 150R_{\rm e}$.  For models B13 and B14, the
simulations were performed in this grid setup.  To confirm the
convergence of numerical results, we also performed a simulation for
the low-resolution model B13L, in which $R_{\rm e}$ is covered by 60
grid points in the finest domain, while keeping the outer boundary at
the same position as the high resolution model.  In these three
models, the equatorial plane symmetry is imposed. We also performed a
simulation for model B14F for which no symmetry is assumed. This
simulation is devoted to exploring the dependence of the symmetry
effect on the evolution of the magnetized neutron stars.

\begin{table*}
\centering
\begin{minipage}{140mm}
\caption{\label{tab:model} Physical parameters of a differentially
rotating neutron star employed: Gravitational mass $M$, baryon rest
mass $M_b$, central density (maximum density) $\rho_{\rm max}$,
angular momentum ${\rm c}J/{\rm G}M^2$, central rotation period $P_c$,
and coordinate radius on the equator $R_{\rm e}$.}
\begin{tabular}{cccccc}
\hline\hline
$M~(M_\odot)$ & $M_b~(M_\odot)$ & $\rho_{\rm max}~({\rm g/cm^3})$ & ${\rm c}J/{\rm G}M^2$ & $P_c~(\rm ms)$ & $R_{\rm e}~({\rm km})$\\
\hline
2.02 & 2.23 & $9.49\times 10^{14}$ & 0.66 & 0.48 & 11.4\\
\hline\hline
\end{tabular}
\end{minipage}
\end{table*}

\begin{table*}
\centering
\begin{minipage}{140mm}
\caption{\label{tab:grid} Model parameters and grid setup: Maximum
strength for the initial dipole magnetic field $B_0$, the finest grid
resolution $\Delta x_{l_{\rm max}}$, the grid point within one
refinement domain $N$, the total number of FMR domains $l_{\rm max}$,
the location of the outer boundary $L_0$ along each axis, and
the assumption for the equatorial plane symmetry. }
\begin{tabular}{ccccccc}
\hline\hline
Model & $B_0$ [G] & $\Delta x_{l_{\rm max}}$ [km] & $N$ & $l_{\rm
max}$ & $L_0$ [km] & eq-symmetry \\
\hline
B14   & $1.7\times 10^{14}$ & 0.142 & 96 & 8 & 1740 & yes \\
B13   & $4.2\times 10^{13}$ & 0.142 & 96 & 8 & 1740 & yes \\
B13L  & $4.2\times 10^{13}$ & 0.190 & 72 & 8 & 1740 & yes \\
B14F  & $1.7\times 10^{14}$ & 0.142 & 96 & 8 & 1740 & no  \\
\hline\hline
\end{tabular}
\end{minipage}
\end{table*}

\section{Numerical results}\label{sec:result}

\subsection{Prediction}

First, we summarize the predicted numerical results based on the
findings of Ref.~\cite{Shibata:2011fj}, and then, describe
three-dimensional effects that are not taken into account in the
previous work~\cite{Shibata:2011fj}.  In the present setup, the
magnetic winding due to the presence of differential rotation and
poloidal magnetic fields will take place and then, a strong toroidal
field will be generated~\footnote{Note that the
MRI~\cite{Balbus:1998ja} may occur in the magnetized neutron
star. However, the wavelength for the fastest growing mode is too
short to be resolved by the present grid resolution:
Assuming the typical value for the HMNS, the MRI wavelength is 
$6 \times 10^3 (B/10^{14}{\rm G}) (\rho/10^{15} {\rm g/cm}^3)^{-1/2} (\Omega / 10^3 {\rm rad/s})^{-1}$~cm. 
Typical resolution of our simulation is an order of $10^{4}$~cm (see Table II). 
If the effect of
the MRI could be incorporated, we may find that the power of the
outflow driven is higher than that found in the present work.}.  The
magnitude of the toroidal magnetic field increases linearly with time
during the winding.  In particular, a strong magnetic field is
generated near the rotation axis.  After the substantial winding, the
magnetic pressure associated with the strong toroidal field overcomes
the gravitational binding energy in the vicinity of the neutron-star
polar surface. Then, a sub- or mildly relativistic outflow will be
launched primarily toward the direction perpendicular to the
equatorial plane.  In the outflow, both a matter outflow and a
Poynting flux are generated.  The magnitudes of
the luminosity for these would be~\cite{Shibata:2011fj}
\begin{align}
 &L_M \sim 10^{48} \left(\frac{B_0}{10^{13}{\rm G}}\right)^2
 \left(\frac{R_{\rm e}}{10^6{\rm cm}}\right)^3\left(\frac{\Omega}{10^4{\rm rad/s}}\right)~{\rm ergs/s},\label{scale1}\\
 &L_B \sim 10^{47} \left(\frac{B_0}{10^{13}{\rm G}}\right)^2
 \left(\frac{R_{\rm e}}{10^6{\rm cm}}\right)^3\left(\frac{\Omega}{10^4{\rm rad/s}}\right)~{\rm ergs/s},\label{scale2}
\end{align}
where $\Omega$ is the typical magnitude of the angular velocity.
Here, $L_M$ includes the contribution of the rest-mass energy flow,
and thus, the luminosity for the kinetic energy would be by about two
orders of magnitude smaller for the outflow velocity $\sim 0.1$ --
0.2 c.

The simulations of Ref.~\cite{Shibata:2011fj} were performed assuming
the axial symmetry. In the nonaxisymmetric case, we should consider
that the nonaxisymmetric kink instability~\cite{Goedbloed} could turn
on because the outflow contains a strong toroidal field generated by
the winding as a dominant magnetic-field component.
Reference~\cite{McKinney:2009} indeed showed that the kink instability
turns on in a magnetically driven jet from a black hole-torus system,
if it has a strong toroidal field. In the following, we will show a
numerical result which illustrates that the kink instability indeed
turns on. The question is how the effect of this instability modifies
the results of the axisymmetric simulations~\cite{Shibata:2011fj}.

\subsection{Properties of outflow}

\begin{figure*}
\begin{center}
\centerline{
\begin{tabular}{cc}
\includegraphics[width=9.5cm,angle=0]{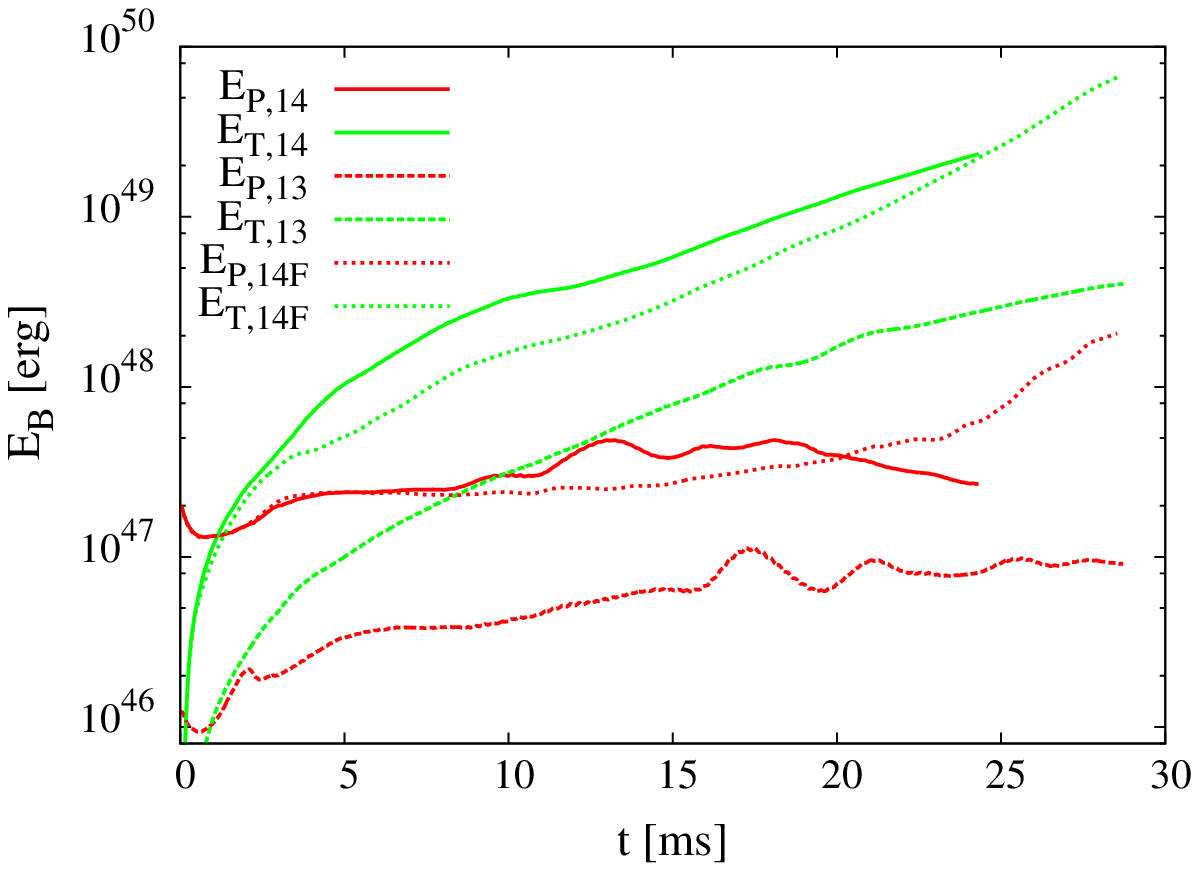} &
\includegraphics[width=9.5cm,angle=0]{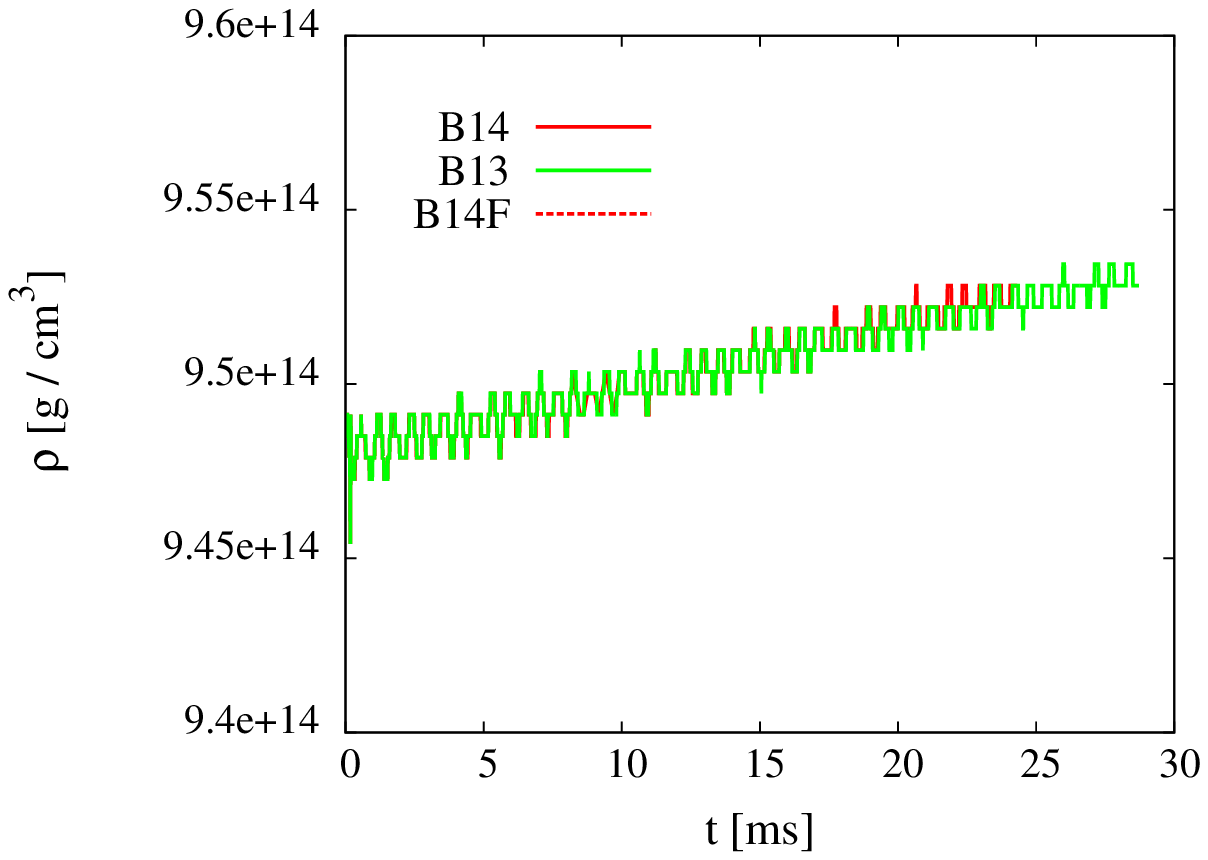}
\end{tabular}
}
\vspace{-3mm}
\caption{\label{fig1} (Left) Evolution of the poloidal-field energy
$E_{\rm P}$ and toroidal-field energy $E_{\rm T}$ for three models
B14, B13, and B14F, labeled by ``14'', ``13'', and ``14F'', respectively.
(Right) Evolution of the central rest-mass density.  }
\end{center}
\end{figure*}

\begin{figure*}
 \begin{center}
   \includegraphics[width=17.0cm,angle=0]{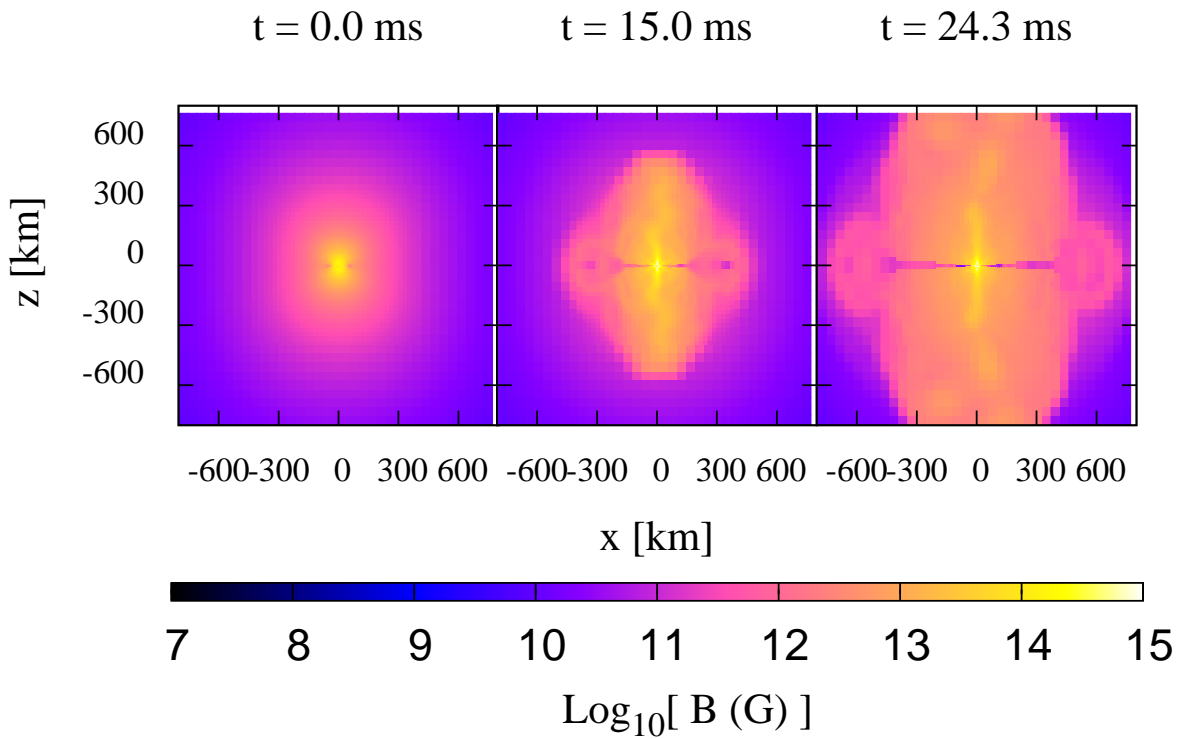}
\vspace{-20mm}
 \caption{\label{fig2} Snapshots of the magnetic field strength on a
 meridional plane ($x$-$z$ plane) at $t=0$~ms (left), $t=15.0$~ms
 (center), and $t=24.3$~ms (right) for model B14.  }
 \end{center}
\end{figure*}

\begin{figure*}
\begin{center}
\includegraphics[width=17.0cm,angle=0]{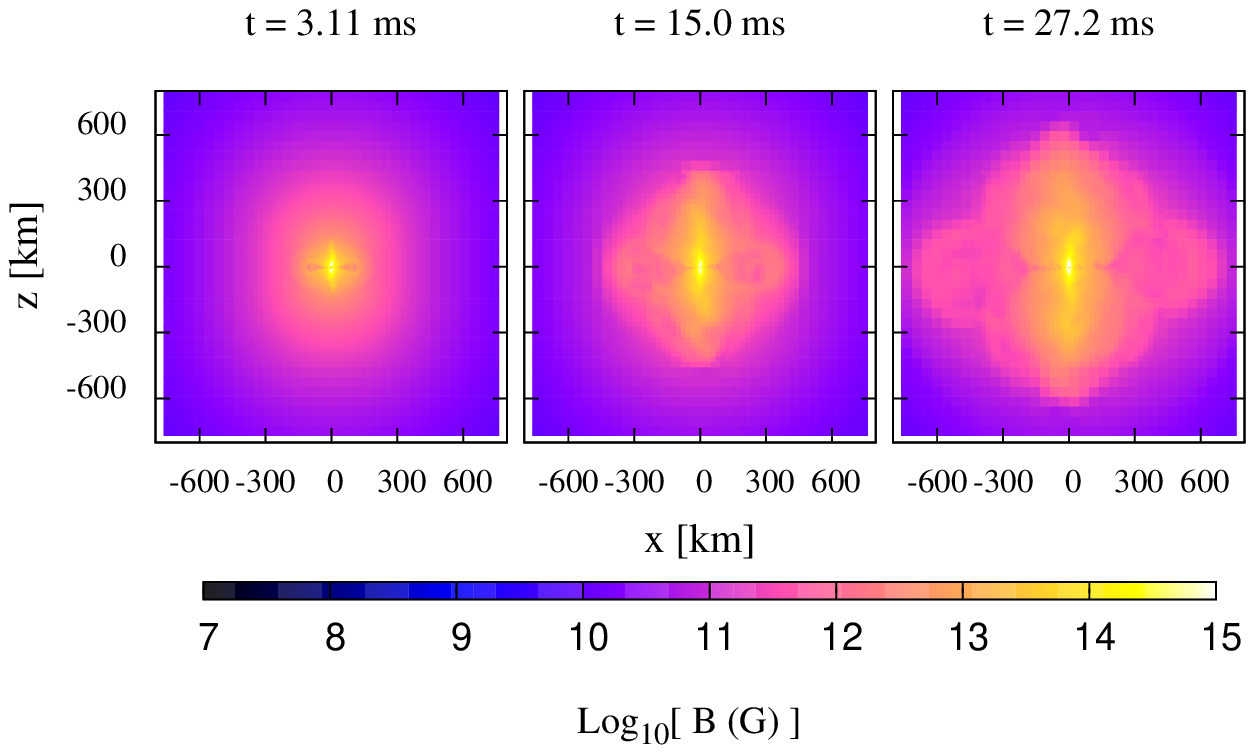}
\vspace{-10mm}
\caption{\label{fig22}
 The same as Fig.~\ref{fig2}, but at $t=3.11$~ms (left), $t=15.0$~ms
 (center), and $t=27.2$~ms (right) for model B14F.  } \end{center}
\end{figure*}

Figure~\ref{fig1} plots the evolution of the electromagnetic energy as
a function of time for all the models employed in this paper. We
define the electromagnetic energy by
\begin{align}
E_B = \frac{1}{8\pi} \int b^2 w \sqrt{\gamma} {\rm d}^3x,
\end{align}
where $\gamma$ is the determinant of the spatial metric and $w=-n_\mu
u^\mu$ with $n^\mu$ being a timelike unit vector normal to the spatial
hypersurface and $u^\mu$ being a four velocity.  $E_B$ is decomposed
into the poloidal component $E_{\rm P}$ and toroidal one $E_{\rm T}$
as
\begin{eqnarray}
E_{\rm P} &=& \frac{1}{8\pi} \int (b^R b_R + b^z b_z) w \sqrt{\gamma}
{\rm d}^3x,\\ E_{\rm T} &=& \frac{1}{8\pi} \int b^\varphi b_\varphi w
\sqrt{\gamma} {\rm d}^3x.
\end{eqnarray}
As expected, the toroidal-field energy for all the models increases
with time due to the magnetic winding.  In a relatively short
timescale $\sim 1$~ms $(\approx 2P_c)$, the toroidal-field energy
catches up with the poloidal one, and then, it becomes the dominant
component.  We find the growth rate of the toroidal field is
consistent with the winding mechanism, in which the toroidal field
$B_{\rm T}$ should increase in proportional to $\sim B_{\rm P} \Omega
t$ with $B_{\rm P}$ being the poloidal magnetic field.  

The toroidal-field energy for model B14F starts deviating from that
for model B14 at $t\sim 3$ ms and the growth rate for model B14F is
slightly smaller than that for model B14. On the other hand, for $t
\agt 24$~ms, the toroidal-field energy for model B14F overcomes that
for model B14.  These facts imply that an asymmetry with respect to
the equatorial plane comes into the play in the amplification process
of the magnetic field (see also the left panel of Fig.~\ref{fig22}),
although this effect does not change the amplification process
qualitatively and significantly.

For all the models, the poloidal-field energy also changes with time,
and is eventually larger than the initial value by a factor of $\sim
2$ -- 10.  If the system is axisymmetric, the poloidal-field energy
changes only by the motion in the meridional plane. Assuming the
conservation of the magnetic flux and mass, the poloidal field
increases in proportional to $\rho^{2/3}$, i.e., by compression (see
e.g., Ref.~\cite{Shibata:2006hr}).  However, as displayed in the
right-panel of Fig.~\ref{fig1}, the central density is approximately
constant during the evolution for all the models. This indicates that
the increase of the poloidal-field energy is not due to the
compression, but to a nonaxisymmetric effect.  We will revisit this
point below. We note that the internal energy is also approximately
constant during the simulations, which also implies that the density
distribution in the inner region of the neutron star is approximately
stationary.  Therefore, the magnetic field evolves passively with
respect to the bulk motion inside the neutron star. This is also
recognized from the left panel of Fig.~\ref{fig1}, which shows
that the profile of the curves of $E_{\rm B}$ for models B13 and B14
is quite similar besides the factor determined by the ratio of the
initial magnetic-field strength.  By contrast, the magnetic field
actively evolves outside the neutron star, as described below. This
active non-linear evolution slightly modifies the scaling relation of
$E_{\rm B}$ that might hold between the models of different initial
magnetic-field strengths.

Figures~\ref{fig2} and \ref{fig22} plot the snapshots of the
magnetic-field strength on a meridional plane ($x$-$z$ plane) at
selected time slices for models B14 and B14F, respectively. The
initially dipole field is distorted by the magnetic winding and
consequently the strong toroidal field is generated near the rotation
axis (see the central part in the middle panels of Figs.~\ref{fig2}
and \ref{fig22}).  Then, an outflow is driven in particular along the
rotation axis (see the middle and right panels of Figs.~\ref{fig2} and
\ref{fig22}).  The outflow keeps blowing for a timescale much longer
than the dynamical one $\sim \rho_{\rm max}^{-1/2}$ and rotation
period $P_c$, because the strong toroidal field continues to be
generated by the differential rotation in the neutron star.  The
asymmetry with respect to the equatorial plane develops for model
B14F, which causes the less efficient winding as shown in
Fig.~\ref{fig1}. This may be also found by comparing the
magnetic-field strength in the equatorial plane between two models;
for model B14, the magnetic-field strength there is weak because of
the symmetry imposed, whereas it is stronger for model B14F. Namely,
the winding occurs less coherently for model B14F.  This less
coherence is likely to stem from the fact that the kink instability
turns on in a stronger way in the absence of the equatorial plane
symmetry (see Sec.~\ref{sec:kink}).  Because of this less coherence,
the toroidal field grows with a longer timescale, and as a result, the
head speed of the outflow for model B14F is slightly slower than that
for model B14 (compare the right panels of Figs.~\ref{fig2}
and~\ref{fig22}).

\begin{figure*}
\begin{center}
\includegraphics[width=12.0cm,angle=0]{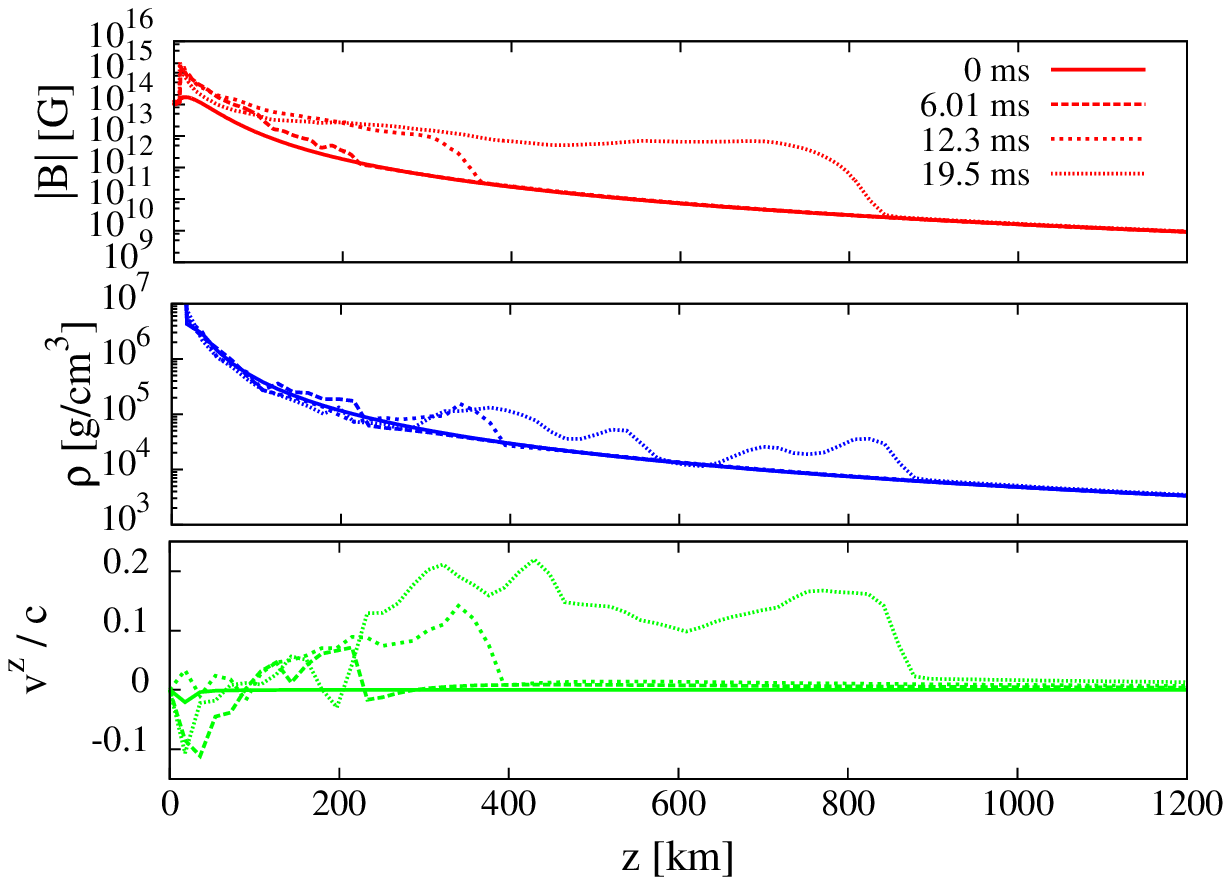}
\caption{\label{fig3} Snapshots for the profiles of the magnetic
field strength (top), rest-mass density (middle), and $z$-component of
the three velocity (bottom) along the $z$-axis at $t=0$ ms (solid curve),
$t=6.01$~ms (dashed curve), $t=12.3$~ms (short-dashed curve), and
$t=19.5$ ms (dotted curve) for model B14.  }
 \end{center}
\end{figure*}

\begin{figure*}
 \begin{center}
   \includegraphics[width=18.0cm,angle=0]{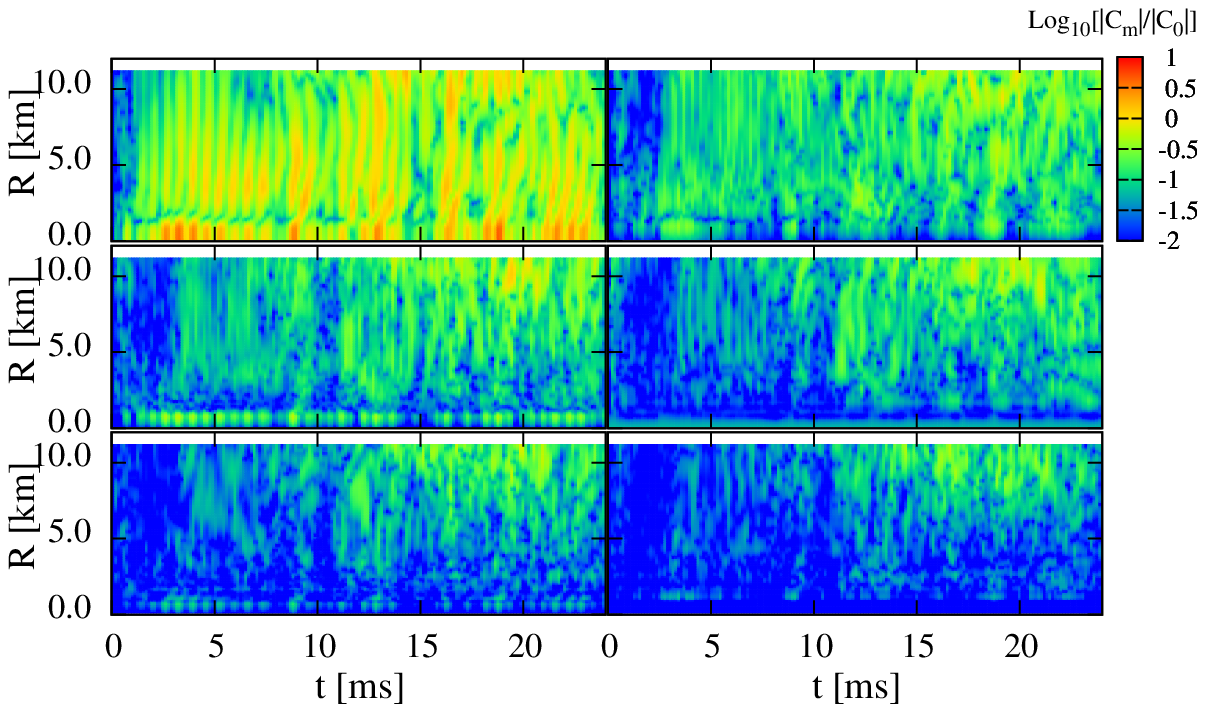}
\vspace{-13mm}
 \caption{\label{fig4a} Fourier mode amplitudes for the toroidal
 field $B_{\rm T}$ as a function of $t$ and cylindrical coordinate $R$
 for model B14 for $m=1$ (left-top), $m=2$ (right-top), $m=3$
 (left-middle), $m=4$ (right-middle), $m=5$ (left-bottom), and $m=6$
 (right-bottom).  }
 \end{center}
\end{figure*}

Figure~\ref{fig3} plots the profiles of the magnetic field, rest-mass
density, and $z$-component of the three velocity along the $z$-axis at
several selected time slices for model B14.  In the vicinity of the
stellar polar surface, the strong magnetic field is generated and its
strength reaches up to $\sim 10^{15}$~G. This substantially winded
magnetic field causes a mass stripping if the magnetic pressure
overcomes the gravitational binding energy density. This is
approximately written by $B^2/8\pi > \rho G M /H$ where $H$ is the
vertical coordinate radius of the neutron star~\cite{Shibata:2011fj}.
This is approximately equivalent to
\begin{equation}
v_{\rm a} > v_{\rm esc}, \label{eq:vave}
\end{equation}
where $v_{\rm a}$ and $v_{\rm esc}$ denote the Alfv{\'e}n speed and
the escape velocity from the stellar surface.  After the substantial
winding, this condition is satisfied for a low-density surface region
of the neutron star, and mass stripping turns on.  Figure~\ref{fig3}
indeed shows that near the stellar surface, $z =H \sim 10$~km, at
$t=6.01$~ms, the Alfv{\'e}n speed $\sim \sqrt{B^2/(4\pi\rho)}$ is of
order the speed of light, c, and hence, Eq.~(\ref{eq:vave}) is
satisfied. After the mass stripping sets in, a blast wave is
generated, and the shock front propagates along the $z$-direction with
its sub-relativistic head speed $\sim 0.1$ -- 0.2~c. We infer that 
differentially rotating magnetized neutron stars will drive the
outflow as far as it is alive. 

\subsection{Kink instability}\label{sec:kink}

A noteworthy new feature that was not able to be found in
Ref.~\cite{Shibata:2011fj} is that a nonaxisymmetric structure of the
magnetic-field profile with respect to the rotation axis emerges in
the outflow; see the middle and right panels in Figs.~\ref{fig2} and
\ref{fig22}.  This implies that a nonaxisymmetric instability sets in.
This nonaxisymmetric profile is found in particular in the vicinity of
the rotation axis.  As already described, the strength of the toroidal
field generated by the magnetic winding is as large as or larger than
the poloidal-field strength in the region near the rotation axis.
It is well known that a cylindrical plasma column surrounded by
toroidal magnetic fields causes the kink instability~\cite{Goedbloed}.
The situation for the vicinity of the rotation axis is quite similar
to the cylindrical plasma column.  This instability is known to turn
on if the following Kruskal-Shafranov (KS) instability criterion is
satisfied~\cite{Goedbloed}:
\begin{align}
\left|\frac{B_{\rm T}}{B_{\rm P}}\right| > \frac{2\pi a}{R_0}. \label{KS}
\end{align}
Here, $a$ and $R_0$ are the radius and poloidal extent of a
cylindrical column, respectively. Because $a$ would be smaller than
$R_{\rm e}$ and $R_0 \approx z $, the condition~(\ref{KS}) can be
easily satisfied once the toroidal-field strength is comparable to the
poloidal-field one.  Hence, we infer that the magnetic outflow driven
from the neutron star is subject to the kink instability.

To determine the dominant mode of the kink instability, we perform a
Fourier mode analysis for the toroidal field by calculating
\begin{align}
C_m(t,R,z_0) \equiv \int B_{\rm T} (t,R,\varphi,z_0) {\rm e}^{-im\varphi}
d\varphi. \label{eq:mode}
\end{align}
Here, the spatial hypersurface is sliced for a sequence of
$z=z_0(=$const) planes on each of which we consider rings of radius
$R$ and perform the azimuthal integral (\ref{eq:mode}) along the
rings. Varying the radius of the rings and selected time, we plot
Fig.~\ref{fig4a} for $z_0\approx 1.9R_{\rm e}$.  This figure shows
that $m=1$ mode turns on in particular in the vicinity of the rotation
axis $R\lesssim 1~{\rm km}$. We find that the ratio $|B_{\rm T}| /
|B_{\rm P}| \approx 1$ at $R\approx 1~{\rm km}$.  The right-hand side
of the KS condition (\ref{KS}) is an order of $0.1$ with $a \sim 1$ km
and $R_0\sim 20$ km. Therefore, the toroidal-field strength comparable
to the poloidal-filed one is large enough to induce the kink
instability.  We find that the modes other than the $m=1$ mode do not
exhibit a remarkable growth. This implies that the $m=1$ mode is the
dominant mode, and it does not cause a strong non-linear mode coupling.

\begin{figure*}
 \begin{center}
 \begin{tabular}{cc}
   \includegraphics[width=8.0cm,angle=0]{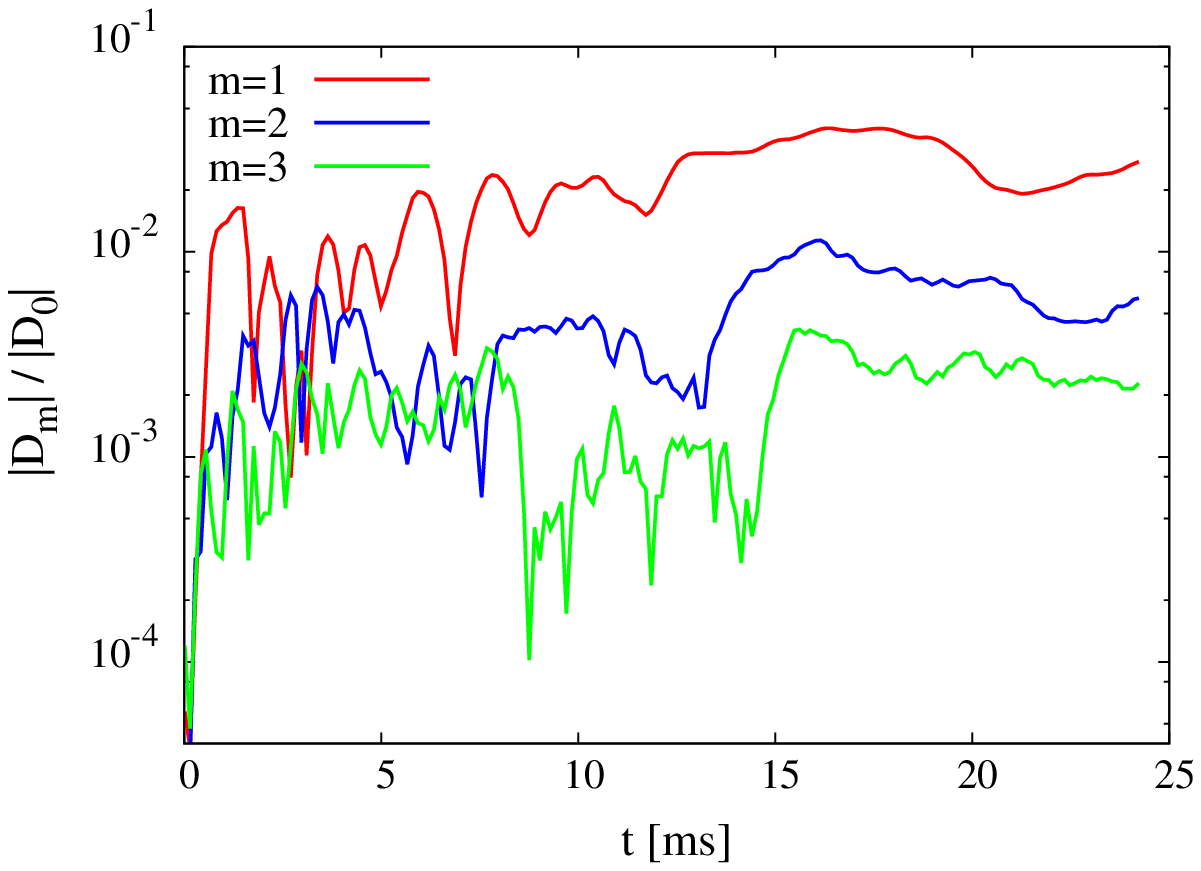} &
   \includegraphics[width=8.0cm,angle=0]{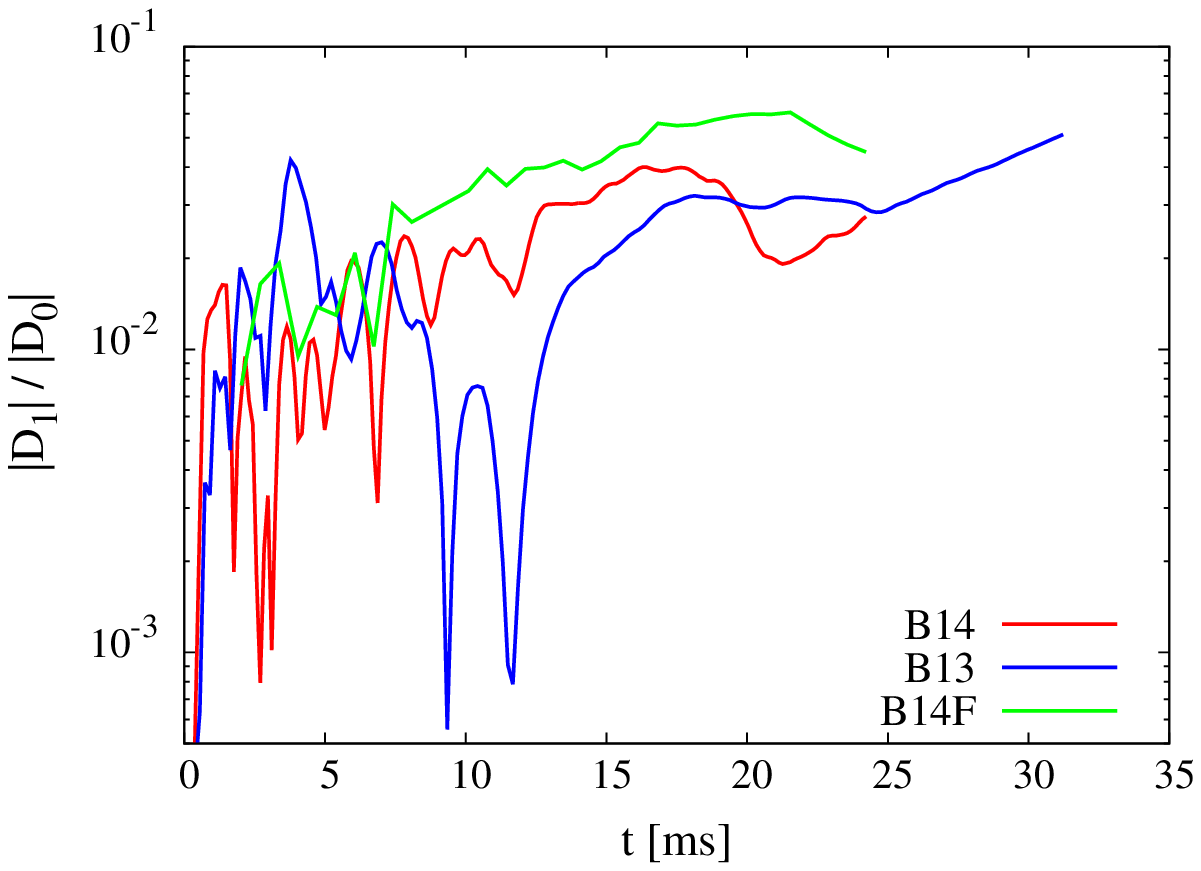}
 \end{tabular}
 \caption{\label{fig4} Evolution of the volume integrated Fourier
 spectrum (a) for $m=1$ -- 3 modes for model B14 and (b) for $m=1$
 mode for models B13, B14, and B14F.  }
 \end{center}
\end{figure*}

\begin{figure*}
 \begin{center}
   \includegraphics[width=10.0cm,angle=0]{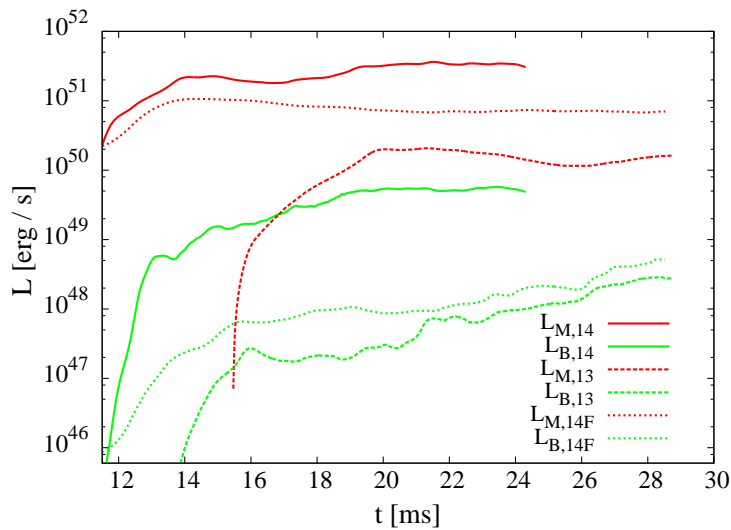}
 \caption{\label{fig5} Evolution of the matter and electromagnetic
 energy ejection rates, $L_M$ and $L_B$, for three models  B14, B13,
 and B14F, labeled by ``14'', ``13'', and ``14F'', respectively. }
 \end{center}
\end{figure*}

\begin{figure*}
 \begin{center}
 \begin{tabular}{cc}
   \includegraphics[width=8.0cm,angle=0]{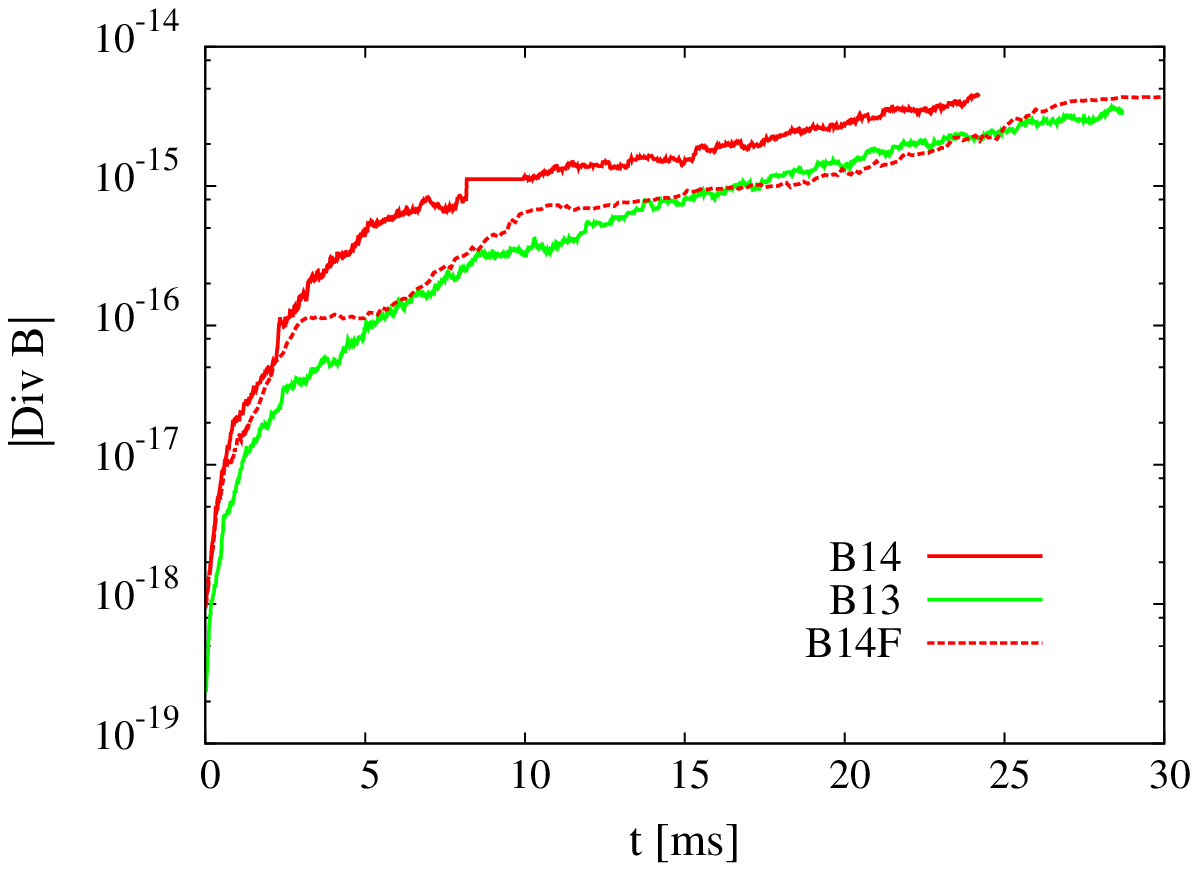} &
   \includegraphics[width=8.0cm,angle=0]{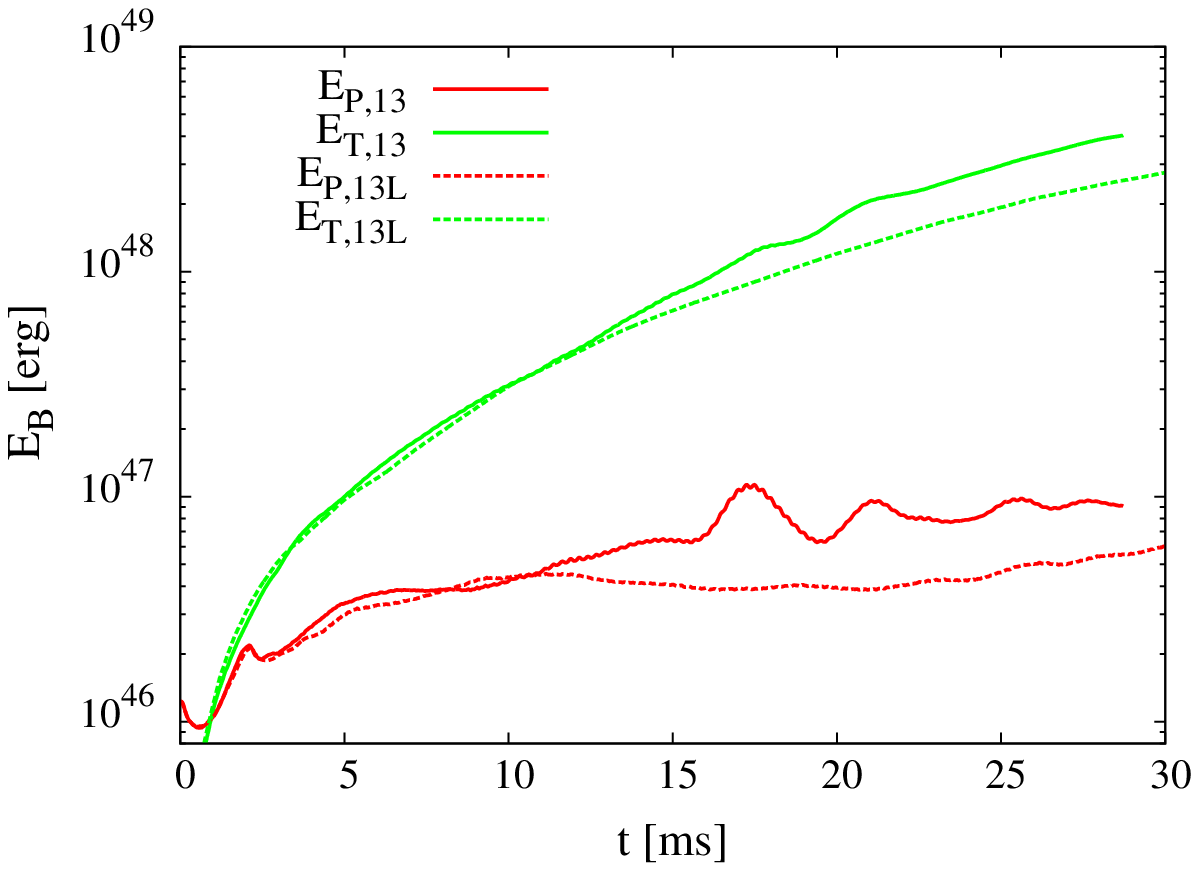}
 \end{tabular}
 \caption{\label{fig8} Evolution of the $L_\infty$ norm of the
 divergence-free condition for all the models (left) and the
 poloidal-field energy $E_{\rm P}$ and toroidal-field energy $E_{\rm
 T}$ as functions of time for models B13 and B13L (right). The right
 panel should be compared with the left panel of Fig.~\ref{fig1}.  }
 \end{center}
\end{figure*}

Figure~\ref{fig4} plots the Fourier components of the toroidal field
as functions of time, which are defined by a volume integral as
\begin{align}
D_m \equiv \int B_{\rm T} {\rm e}^{-im\varphi} d^3 x.
\end{align}
The left panel of Fig.~\ref{fig4} plots the evolution of $D_m$ for
model B14.  This shows that the $m=1$ mode is dominant among the $m=1$
-- $3$ modes as expected from Fig.~\ref{fig4a}, and its saturation
amplitude is at most $\approx~2$ -- 3 percents of the axisymmetric
mode. This again clarifies that the non-linearity of the kink
instability is not strong enough to modify the outflow structure
significantly. It should be pointed out that this weak non-linearity
was also found for the magnetic jet driven from the black hole-torus
system in the simulation of Ref.~\cite{McKinney:2009}. The features
found for model B14 also hold for other models: The right panel of
Fig.~\ref{fig4} shows the evolution of the $m=1$ mode for three
models, showing the weak growth of the $m=1$ mode.

One point to be noted is that the saturation amplitude for model B14F
is slightly larger than that of other models. The likely reason is
that the absence of the equatorial-plane symmetry would enhance the
growth of the kink instability (in other words, the presence of
this symmetry would suppress the growth for some channel of the kink
instability). The stronger excitation of the kink instability for
model B14F results in a stronger modification of the axisymmetric
outflow structure, which we found in Fig.~\ref{fig22}.

\subsection{Luminosities}

The magnetic outflow driven from the magnetized neutron star is
accompanied by a large amount of the ejected material and
electromagnetic waves.  We here define the luminosities for them by
\begin{align}
& L_M = - \oint_{r=\rm const.} d\theta d\varphi \sqrt{-g}
{(T^{\rm (mat)})^r}_t,\\
& L_B = - \oint_{r=\rm const.} d\theta d\varphi \sqrt{-g}
{(T^{\rm (em)})^r}_t,
\end{align}
where $T^{\rm (mat)}_{\mu\nu}$ and $T^{\rm (em)}_{\mu\nu}$ are the
stress energy tensor for the matter and electromagnetic field,
respectively.  $g$ is the determinant of the spacetime metric.  We
note again that $L_M$ includes the contribution of the rest-mass
energy flow and the contribution only from the kinetic energy is by
about two orders of magnitude smaller than $L_M$ for the outflow
velocity $\sim 0.1$ -- 0.2 c.

Figure~\ref{fig5} displays the evolution of these luminosities, which
are calculated for the extraction radius of $r_{\rm ex} \approx
420$~km. We note that the extraction was performed for several radii
and we confirmed that their luminosities depend only weakly on the
extraction radii.  Comparing Figs.~\ref{fig3} and \ref{fig5}, we find
that for model B14, the outflow front reaches the extraction point at
$t\approx 12$ ms, which corresponds to the moment of a quick rise of
the luminosities. Subsequently, the order of the magnitude of the
luminosities remains approximately unchanged.  For other models, the
feature of the luminosity curves is essentially the same.  The matter
energy flux for model B14 attains an order of $10^{51}$~erg/s.
Figure~\ref{fig3} shows that the outflow density at the extraction
point for this model is approximately $10^{5}~{\rm g/cm^3}$ and $v^z
\approx 0.1 c$. These values are consistent with the matter energy
flux if we assume that the matter is ejected quasispherically, i.e.,
$L_M \sim 4 \pi r_{\rm ex}^2 \rho c^2 v^z$.  For the electromagnetic
radiation, the luminosities for models B13 and B14 are consistent with
the scaling relation~(\ref{scale2}), which implies that the scenario
described in Ref.~\cite{Shibata:2011fj} is still valid even in the
presence of the kink instability.  On the other hand, the
electromagnetic luminosity for model B14F is about 10 times smaller
than that for model B14 at the end of the simulation.  This is because
the magnetic winding occurs less coherently due to a stronger effect
of the kink instability, as already discussed in
Sec. \ref{sec:kink}. However, during the longterm evolution, the
toroidal-field energy for model B14F surpasses that for model B14 (see
Fig.~\ref{fig1}) and in addition, Fig.~\ref{fig5} shows that the
electromagnetic luminosity for model B14F increases gradually with
time.  This suggests that although the increase timescale of the
toroidal-field energy and the electromagnetic luminosity is rather
long, the luminosity for model B14F may be eventually as large as that
for model B14.

We find that the scaling relation for the matter flux~(\ref{scale1})
also holds. We confirm that our result basically agrees with that in
Ref.~\cite{Shibata:2011fj}. This is because the saturation amplitude
of the kink instability is not large enough to disrupt the coherent
toroidal magnetic-field profile, as already discussed.

\subsection{Accuracy check}


Finally, we comment on the reliability of the present simulation
results. The left panel of Fig.~\ref{fig8} plots the magnitude of the
violation for the divergence-free condition and the convergence for
the evolution of $E_{\rm B}$ as functions of time. We plot the
$L_\infty$ norm of $\nabla \cdot {\bf B}$ for models B14, B13, and
B14F. The divergence-free condition is well satisfied throughout the
simulations, which means that our FMR implementation for the magnetic
field works well.

The right panel of Fig.~\ref{fig8} plots the evolution of the
magnetic-field energy for model B13 with two grid resolutions.  The
magnetic-field energy for both runs sufficiently converge until
$t\approx 10$~ms and the poloidal-field energy starts deviating after
that, although the behavior of the evolution is qualitatively the
same.  We attribute this loss of the convergence to the longterm
accumulation of the numerical errors such as spurious magnetic
reconnections due to the poor resolution.  We confirmed that the
qualitative features of our finding, e.g., the emergence of the kink
instability and the generation of the outflow, are not affected by the
grid resolution. Therefore, we conclude that the present grid setup is
fine enough for obtaining scientific results for the evolution of
differentially rotating magnetized neutron stars.

\section{Discussion and summary}\label{sec:summary}

\subsection{Discussion}

We here discuss possible electromagnetic signals emitted by the ejecta
from a HMNS formed after the merger of BNS, referring to the numerical
results in the present work.  As mentioned in Sec.~I, the recent
observational result of PSR J1614-2230 suggests that the maximum mass
of spherical neutron stars should be larger than $1.97 \pm 0.04
M_\odot$.  This indicates that the EOS of neutron stars is stiff, and
thus, a long-lived HMNS would be a canonical outcome of the BNS
merger, if the binaries are composed of neutron stars of a canonical
mass of $1.3$ -- $1.4M_\odot$ with the total mass $\sim
2.7M_{\odot}$~\cite{Hotokezaka:2011dh}.

Electromagnetic signals should be emitted from the ejected material of
sub-relativistic motion or ejected electromagnetic waves. According to
recent studies~\cite{Nakar:2011,Metzger:2010,MB2012} the ejected
material will sweep up the interstellar matter and form blast
waves. During this process turning on~\cite{Nakar:2011}, the shocked
material could generate magnetic fields and accelerate particles that
emit synchrotron radiation, for a hypothetical amplification of the
electromagnetic field and a hypothetical electron injection. The
emission will peak when the total swept-up mass approaches the ejected
mass, because the blast wave begins to decelerate according to a
Sedov-Taylor's self-similar solution.  The predicted deceleration time
depends on the total energy $E_0$ and speed of the ejected material
$\beta_0$c as well as the number density of the interstellar matter
$n_0$ for a single velocity outflow as~\cite{Nakar:2011}
\begin{equation}
\sim 2~{\rm yrs} \biggl({E_0 \over 10^{49}~{\rm erg}}\biggr)^{1/3}
 \biggl({n_0 \over 1~{\rm cm}^{-3}}\biggr)^{-1/3} \biggl({\beta_0 \over
 0.2}\biggr)^{-5/3} .
\end{equation}
Here, the value of $n_0$ will depend strongly on the site where the
merger of BNS happens. If the site is in a galactic disk, $n_0$ would be
$\sim 1~{\rm cm}^{-3}$, whereas if it is outside a galaxy, the value is
much smaller $\sim 10^{-3}~{\rm cm}^{-3}$.  By the synchrotron
radiation, a radio signal of $\sim 0.1$ GHz, which is determined by the
self-absorption, could be emitted as in the afterglow of gamma-ray
bursts~\cite{Nakar:2011} for $n_0 \sim 1~{\rm cm}^{-3}$ and $\beta=0.2$.
Then, its luminosity and flux would depend on the total kinetic
energy. Figure~\ref{fig5} indicates that for the model with the
initial maximum field strength of $10^{14}$~G, the luminosity of the
total matter energy (including the rest mass, internal, and kinetic
energies) is $\sim 10^{50}$ -- $10^{51}$~erg/s. Because the typical
speed of the ejected material is sub-relativistic with $\beta_0 \sim
0.1$ -- 0.2, the luminosity of the kinetic energy would be $\sim
10^{48}$ -- $10^{49}$~erg/s. The latest numerical-relativity
simulations indicate that the lifetime of the long-lived HMNS would be
0.1 -- 1~s~\cite{Hotokezaka:2011dh,Sekiguchi:2011zd}. Thus, the
predicted total kinetic energy ejected will be at most $\sim
10^{49}$~erg for $B_0=10^{14}$~G. If the magnetic-field strength is
smaller than $10^{14}$~G, this value is smaller by a factor
$(B_0/10^{14}~{\rm G})^2$. The unabsorbed flux at the typical
synchrotron frequency is
\begin{align}
 \sim & 2.5~{\rm mJy} \left( \frac{E_0}{10^{49}~{\rm erg}} \right)
 \left( \frac{n_0}{1~{\rm cm}^{-3}} \right)^{1/2} \left(
 \frac{\beta_0}{0.2} \right)^{-1} \nonumber \\
 & \times \left( \frac{D}{300~{\rm Mpc}}\right)^{-2} ,
\end{align}
and the peak flux at the self-absorption frequency is approximately two
orders-of-magnitude smaller, i.e., $O(10) \mu {\rm Jy}$ level. The
studies of Ref.~\cite{Nakar:2011} (see Table 1 of it) suggest that the
total energy of $10^{49}$~erg for the sub-relativistic outflow is not
large enough to produce a radio signal observable by current and planned
radio telescopes even for the optimistic value $n_0=1~{\rm cm}^{-3}$
($\sim 10^{50}$~erg is suggested to be necessary), and our estimate
agrees with their results. Thus, this mass-ejection mechanism is
unlikely to supply a large amount of the mass which generates a
sufficiently strong radio signal, unless the magnetic-field strength is
extremely large, as large as that of magnetars for which the field
strength could be $\sim 10^{15}$~G.

Alternatively, the authors of Ref.~\cite{Metzger:2010,MB2012,Kulkarni} (see
also Ref.~\cite{LP1998} for the original idea) discuss the signals by
the radioactive decay of $r$-process nuclei, which are produced from the
neutron-rich material in the outflow, and subsequently decay and emit
a signal that may be observable by current and planned optical
telescopes such as LSST. In this scenario, the typical duration of the
peak luminosity is of order a day or less as \cite{LP1998}
\begin{eqnarray}
t_{\rm peak} \approx 0.1{\rm d}\biggl({\beta_0 \over 0.2}\biggr)^{-1/2}
\biggl({M_{*,{\rm esc}} \over 10^{-3}M_{\odot}}\biggr)^{1/2},
\end{eqnarray}
and the associated peak luminosity is
\begin{eqnarray}
L_{\rm peak} &\approx& 7 \times 10^{41}~{\rm erg/s}
\biggl({f \over 3 \times 10^{-6}}\biggr) \nonumber \\
&&\times \biggl({\beta_0 \over 0.2}\biggr)^{1/2}
\biggl({M_{*,{\rm esc}} \over 10^{-3}M_{\odot}}\biggr)^{1/2},
\end{eqnarray}
where $M_{*,{\rm esc}}$ is the total amount of the rest mass ejected and 
$f$ denotes the conversion rate of the energy per rest-mass
energy in the ejected material through the radioactive decay process,
which is $\sim 3 \times 10^{-6}$ according to the results
of~\cite{Metzger:2010}. According to Ref.~\cite{Metzger:2010,MB2012},
if the total ejected mass is $\agt 10^{-3}M_{\odot}$, the signal
can be detected by large optical surveys. Figure~\ref{fig5} indicates
that the mass ejection rate is $\sim 10^{-4}$ -- $10^{-3}M_{\odot}$/s
for the maximum field strength of $10^{14}$~G. Thus, even if the
lifetime of the HMNS is 1~s, total amount of the rest mass ejected
will be $\sim 10^{-4}$ -- $10^{-3}M_{\odot}$ for this field
strength. Again, unless the magnetic-field strength of the HMNS is
extremely large (as large as that of magnetars), the HMNS will not
eject the material which subsequently can be detected by current and
planned optical telescopes.

It should be noted that our estimation is based on the magnetically
driven outflow.  BNS will eject a large amount of mass ($10^{-3}$ --
$10^{-2}M_{\odot}$) with the velocity 0.2 -- 0.3c through the
dynamical torque that works during the merger
process~\cite{H2012}. Such a material will be also composed primarily
of neutrons and produce an amount of unstable $r$-process nuclei, which
subsequently decay and emit a signal that may be observable by current
and planned optical telescopes~\cite{Metzger:2010}.  The large amount
of the ejected materials may also contribute to generate radio signals
interacting with the interstellar matter, as argued in
Ref.~\cite{Nakar:2011}.


Most important finding in the previous~\cite{Shibata:2011fj} and
present works is that electromagnetic waves are emitted from the HMNS
together with the mass ejection. This implies that even in the absence
of the generation of blast waves via the interaction with the
interstellar material, a strong magnetic field is generated. We find
that the electromagnetic luminosity is $10^{49}$ -- $10^{50}$~erg/s
for the model of the maximum field strength of $10^{14}$~G. For the
hypothetical lifetime of the HMNS of 0.1 -- 1~s, the total radiated
energy by electromagnetic waves is $\sim 10^{49}$~erg, which is larger
than the total kinetic energy of the ejected material in our
model. Such huge magnetic energy, composed primarily of Alfv{\'e}n
waves, may be reprocessed efficiently to an observable signal as in
the solar corona, although the mechanism is not clear.  To clarify
this point, a first-principle simulation taking into account detailed
physical processes, as done, e.g., in Ref.~\cite{Suzuki} for the solar
corona problem, will be necessary.

Finally, we should comment on the saturation of magnetic field strength. 
As shown in 
Fig.~\ref{fig1}, the magnetic field continues to grow at the end of the simulation for all the models. 
The magnetic field strength would saturate if the magnetic field energy is as large as the kinetic energy or the 
thermal energy of the HMNS. 
We estimate the kinetic energy as $\sim 10^{53}$ erg and growth rate of the magnetic field energy in Fig.~\ref{fig1} as 
$\sim 10^{49}~{\rm erg}/{\rm ms}$, where we assume the toroidal magnetic field energy continues to grow by the magnetic winding 
and use model B14F as a representative model. Then, the magnetic field would saturate at $t\sim 100$ ms, which is shorter than 
the lifetime of the HMNS $\sim 1$~s. Therefore, our scaling relation~(\ref{scale1})--(\ref{scale2}) would breakdown after the saturation 
and observational signature might be altered after that. On the other hand, if the initial magnetic field of HMNS is not strong, e.g., 
$10^{13}$ G, or the HMNS lifetime is O(0.1)s, the magnetic field strength continues to increase and the HMNS would collapse to a BH before the magnetic field 
saturate and Eqs.~(\ref{scale1})--(\ref{scale2}) would hold. 

\subsection{Summary}

Using a new GRMHD code implementing the FMR algorithm based on the
divergence-free interpolation scheme of Balsara~\cite{Balsara:2001rw},
we performed numerical simulations for the evolution of a
differentially rotating magnetized neutron star, as an extension of
our previous axisymmetric work~\cite{Shibata:2011fj}.  The magnetic
winding mechanism generates the strong toroidal field in particular in
the vicinity of the rotation axis, as in the axisymmetric
case. Subsequently, Alfv\'{e}n waves propagate primarily toward the
$z$-direction along the rotation axis and transport the
electromagnetic energy.  After substantial winding, the magnetic
pressure overcomes the gravitational binding energy density and drives
a sub-relativistic outflow.  We found that in this tower-type outflow,
the kink instability develops due to the presence of a strong toroidal
magnetic field, and modifies the structure of the outflow that would
have an axisymmetric structure in the absence of this instability.
However, this instability saturates in a relatively small amplitude
level, and thus, does not significantly modify the profile of the
outflow. We also confirmed that the scaling relations for the matter
and Poynting luminosities~(\ref{scale1})--(\ref{scale2}), originally
found in Ref.~\cite{Shibata:2011fj}, hold in the nonaxisymmetric
situation as well.

As mentioned for several times in this paper, the recent observation
of PSR J1614-2230 suggests that the maximum mass of spherical neutron
star has to be larger than $1.97 \pm 0.04 M_\odot$, and implies that a
long-lived HMNS is likely to be a canonical product of the BNS merger
if the binaries are composed of neutron stars with a canonical mass of
$1.3$ -- $1.4M_\odot$~\cite{Hotokezaka:2011dh}. In the formed HMNS, a
magnetic field will be amplified not only by the magnetic winding but
also by the MRI. The MRI could cause an efficient angular momentum
transport. If the strong self-gravity of a HMNS is supported primarily
by its rapid rotation, the angular momentum transport could induce the
collapse of the HMNS to a black hole surrounded by an accretion
torus. (This is not the case if a HMNS is supported mainly by a
thermal pressure~\cite{Sekiguchi:2011zd}.) The black hole-torus system
formed in this scenario is a promising candidate of a central engine
of short gamma-ray
bursts~\cite{Narayan,Duez:2005cj,Shibata:2005mz,Rezzolla:2011da}.  We
also plan to explore this scenario in the future.  The most
self-consistent approach for the study of these scenarios is to
simulate the merger of magnetized BNS taking into
account a plausible EOS for a long time from the late
inspiral to the longterm evolution of the formed HMNS. We also plan to
perform this type of the simulations in the future.

\acknowledgments

We thank K. Hotokezaka, Y. Sekiguchi, and Y. Suwa for helpful
discussions about the possible signals emitted from the outflow.
Kiuchi also thanks T. Muranushi for a valuable comment.  Numerical
simulations were performed on SR16000 at YITP of Kyoto University, and
XT4 at CfCA of NAOJ.  This work was supported by the Grant-in-Aid for
Scientific Research (21340051, 21684014, 24244028, 24740163),
Scientific Research on Innovative Area (20105004), and HPCI Stragetic
Program of Japanese MEXT.

\begin{appendix}

\newcommand{\bB}{{\mbox{\boldmath$B$}}}
\newcommand{\bE}{{\mbox{\boldmath$E$}}}

\section{Code description}

In this appendix, we describe our integration scheme for the induction
equation and our implementation of the FMR algorithm in details.
Following Ref.~\cite{Shibata:2005gp}, we choose the three magnetic
field $\bB^\mu \equiv \sqrt{\gamma}n_\nu F^{*\nu\mu}$ as the basic
variable where $F^{*\nu\mu}$ is the dual of the Faraday tensor.  This
magnetic field is purely spatial in the sense that $n_\mu
\bB^\mu=0$. Assuming that the ideal MHD condition holds, Maxwell's
equation is recasted into the divergence-free condition and induction
equation:
 \begin{align}
 & \partial_a \bB^a = 0 \label{eq:non-mono}\\
 & \partial_t \bB^a = \partial_b [(\bB^b v^a-\bB^a v^b)],
 \label{eq:induct}
 \end{align}
where $v^a=u^a/u^0$. We define the corresponding electric field by
 \begin{align}
& \bE_x=-v^y \bB^z + v^z \bB^y \nonumber\\ & \bE_y=-v^z
 \bB^x + v^x \bB^z \nonumber\\ & \bE_z=-v^x \bB^y + v^y \bB^x.
 \end{align}
Here, the electric field is related to the flux of the induction
equation by $-\bB^b v^a+\bB^a v^b=e^{abc}\bE_c$ with $e^{abc}$ being
the completely antisymmetric symbol in the flat space satisfying
$e^{xyz}=1$.  In the following, we do not distinguish $\bE^a$ and
$\bE_a$.

\subsection{Staggered cell}\label{sec:FMRgrid}

The most popular and robust finite volume method to integrate ideal
MHD equations is the constrained-transport (CT) scheme~\cite{Evans:1988qd}.
In this scheme, a cell for the numerical computation is defined so
that fluid variables are placed in the cell center.  Then, the
(surface averaged) magnetic field and electric field, which is
calculated from the magnetic field and velocity field, are placed on
the cell surfaces and cell edges, respectively, to preserve the
divergence-free condition during the evolution of the magnetic field.
This prescription is compatible with an AMR or FMR implementation of
Balsara~\cite{Balsara:2001rw}, if the cell-centered grid is employed.
For this case, the cell surface of a parent domain always agrees with
the cell surfaces of its child domains, and it becomes straightforward
to guarantee that the magnetic flux penetrating a surface in a
parent-domain's cell agrees with the sum of the magnetic fluxes
penetrating the corresponding surfaces in children-domain's cells.

However, as mentioned in Sec.~\ref{sec:form}, our code is designed for
the vertex-centered grid, because it is suited for integrating
Einstein's equations. This implies that the magnetic field should be
placed at each cell in a different way so that surfaces of defining
the magnetic field in a parent domain agree with surfaces of its child
domains. Specifically, we define cells and place the magnetic field,
as Fig.~\ref{fig:cellB} shows. Because the vertex-centered grid is
employed, the geometric and fluid variables are placed at the cell
corner (i.e., at the grid). We label each grid by $(i,j,k)$ for the
child domain and by $(I,J,K)$ for the parent domain. Then, the
magnetic field is placed on the cell surface, labeled by $(i\pm
1/2,j\pm 1/2,k)$, $(i,j\pm 1/2,k\pm 1/2)$, and $(i\pm 1/2,j,k\pm 1/2)$
for the child cell. The magnetic field is defined in the range of $[-(N+1/2)\Delta x_l, (N+1/2)\Delta x_l]$. 
The CT scheme requires the flux (electric field) to be placed at the
cell edge, labeled by $(i\pm 1/2,j,k)$, $(i,j\pm 1/2,k)$, and
$(i,j,k\pm 1/2)$ for the child cell.  In the following, big letters
such as $B^x$ and $E^x$ denote the components of $\bB^a$ and $\bE_a$
in the parent domain, and small letters such as $b^x$ and $e^x$ do
those in the child domain.

The parent cell contains the eight child cells and the surfaces of the
parent cells always overlap with some cell surfaces of the child
domain. This grid structure is essential for guaranteeing the
divergence-free condition and the magnetic-flux conservation.
Table~\ref{tab:BandE} summarizes the locations where the magnetic and
electric fields are defined. Then, Eq.~(\ref{eq:induct}) is
discretized straightforwardly as
\begin{align}
 \partial_t (B^x)_{I,J+\frac{1}{2},K+\frac{1}{2}}=&
- \frac{(E^z)_{I,J+1,K+\frac{1}{2}}
-(E^z)_{I,J,K+\frac{1}{2}}}{\Delta y_l} \nonumber\\
& + \frac{(E^y)_{I,J+\frac{1}{2},K+1}
-(E^y)_{I,J+\frac{1}{2},K}}{\Delta z_l}, \nonumber\\
 \partial_t (B^y)_{I+\frac{1}{2},J,K+\frac{1}{2}}=&
- \frac{(E^x)_{I+\frac{1}{2},J,K+1}
-(E^x)_{I+\frac{1}{2},J,K}}{\Delta z_l} \nonumber\\
& + \frac{(E^z)_{I+1,J,K+\frac{1}{2}}
-(E^z)_{I,J,K+\frac{1}{2}}}{\Delta x_l}, \nonumber\\
 \partial_t (B^z)_{I+\frac{1}{2},J+\frac{1}{2},K}=&
- \frac{(E^y)_{I+1,J+\frac{1}{2},K}
-(E^y)_{I,J+\frac{1}{2},K}}{\Delta x_l} \nonumber\\
& + \frac{(E^x)_{I+\frac{1}{2},J+1,K}
-(E^x)_{I+\frac{1}{2},J,K}}{\Delta y_l},
\end{align}
where $(\Delta x_l, \Delta y_l, \Delta z_l)$ denote the grid spacing
for $(x,y,z)$ in the parent level labeled by $l$.

According to Refs.~\cite{Delzanna:2003,Shibata:2005gp}, the electric
field is computed at the cell edge by the Lax-Friedrichs formula,
given by
\begin{align}
E^x =& \frac{(E^x)^{LL}+(E^x)^{LR}+(E^x)^{RL}+(E^x)^{RR}}{4}
\nonumber\\
&+\frac{c_y}{2} (B^z_R-B^z_L) - \frac{c_z}{2} (B^y_R-B^y_L) \label{eq:Ex}
\end{align}
at $(I+1/2,J,K)$. 
To evaluate the flux, the magnetic field defined on the cell surface should be interpolated to 
the cell edge where the electric field is defined (see Fig.~\ref{fig:cellB}). 
This means that the magnetic field has the reconstructed right and left state. 
According to the prescription of the central scheme~\cite{Kruganov:2000}, 
we need an offset based on the characteristic speed in the flux. 
In the above equation, $(E^x)^{RL}$ represents the
reconstructed right state in the $y$-direction and left state in the
$z$-direction. The other symbols $(E^x)^{LL}$, $(E^x)^{LR}$, and
$(E^x)^{RR}$ are interpreted in the similar way.  $(B^y_R,
B^y_L)$ and $(B^z_R, B^z_L)$ also denote the right and left
state of $\bB^y$ and $\bB^z$ reconstructed.  $c_y$ and $c_z$ are the
characteristic speeds in the prescription of the upwind flux
construction and calculated at cell edges with the interpolated
variables. We simply calculate these quantities by averaging:
\begin{align}
& (c_y)_{I+\frac{1}{2},J,K}=\frac{(v^y)_{I,J,K}+(v^y)_{I+1,J,K}}{2} \nonumber\\
& (c_z)_{I+\frac{1}{2},J,K}=\frac{(v^z)_{I,J,K}+(v^z)_{I+1,J,K}}{2}.
\end{align}
The formula for $E^y$ ($E^z$) is obtained from Eq.~(\ref{eq:Ex})
by the permutation of the indices $x \to y$, $y \to z$, and $z \to x $
($x\to z$, $y \to x$, and $z \to y$).

For solving GRMHD equations and Einstein's equations, one needs the
magnetic field defined at $(I,J,K)$.  This is done by a simple
averaging as
\begin{align}
 (B^x)_{I,J,K}=& \frac{1}{4}\Big[(B^x)_{I,J+\frac{1}{2},K+\frac{1}{2}}
+ (B^x)_{I,J-\frac{1}{2},K+\frac{1}{2}} \nonumber\\
&+ (B^x)_{I,J+\frac{1}{2},K-\frac{1}{2}}
+ (B^x)_{I,J-\frac{1}{2},K-\frac{1}{2}}\Big] \nonumber\\
 (B^y)_{I,J,K}=& \frac{1}{4}\Big[(B^y)_{I+\frac{1}{2},J,K+\frac{1}{2}}
+ (B^y)_{I+\frac{1}{2},J,K-\frac{1}{2}} \nonumber\\
&+ (B^y)_{I-\frac{1}{2},J,K+\frac{1}{2}}
+ (B^y)_{I-\frac{1}{2},J,K-\frac{1}{2}}\Big] \nonumber\\
 (B^z)_{I,J,K}=& \frac{1}{4}\Big[(B^z)_{I+\frac{1}{2},J+\frac{1}{2},I}
+ (B^z)_{I-\frac{1}{2},J+\frac{1}{2},K} \nonumber\\
&+ (B^z)_{I+\frac{1}{2},J-\frac{1}{2},K}
+ (B^z)_{I-\frac{1}{2},J-\frac{1}{2},K}\Big].
\end{align}

\subsection{FMR implementation}

For the assignment method of the variables in the interior of a cell
described in Sec.~\ref{sec:FMRgrid}, we exploit the divergence-free
reconstruction scheme in the refinement boundaries of the FMR
algorithm, following Refs.~\cite{Balsara:2001rw,Balsara:2009}. First,
we review how to reconstruct the magnetic field in the whole region of
a cell in this scheme.

Consider a cell defined for a domain composed of $x \in
[x_{I},x_{I+1}]$, $y \in [y_{J},y_{J+1}]$, and $z \in [z_{K},z_{K+1}]$
at a FMR level $l$.  In the first step, the magnetic-field profile
on the cell surfaces are reconstructed. When we design a third-order
accurate code for the spatial direction, the profile of the magnetic
field, say $\tilde{B}^x$, on the surface at $x=x_I, y\in
[y_{J},y_{J+1}], z \in [z_{K},z_{K+1}]$ should be written as
\begin{align}
&\tilde{B}^x(x=x_I,y,z)=B^x_0+B^x_y P_1(y)+B^x_z P_1(z) \nonumber\\
&~~~+ B^x_{yy} P_2(y) + B^x_{yz} P_1(y)P_1(z) + B^x_{zz} P_2(z),
\label{eq:surf}
\end{align}
where $P_1(y)=y-y_{J+1/2}$ and $P_2(y)=(y-y_{J+1/2})^2-\Delta
y_l^2/12$. We employ the WENO5 scheme to
obtain the coefficients $B^x_0$, $B^x_y$, $B^x_z$, $B^x_{yy}$,
$B^x_{yz},$ and $B^x_{zz}$~\cite{Balsara:2009}.
In this scheme, we first consider the one-dimensional reconstruction
problem in a zone centered at $y=y_{J+1/2}$, taking into account five
neighboring variables
$\{(B^x)_{J-3/2},(B^x)_{J-1/2},(B^x)_{J+1/2},(B^x)_{J+3/2},(B^x)_{J+5/2}\}$,
where we omit the index $I$ and $K$. Then,
a third-order reconstruction over the zone centered at $y_{J+1/2}$ can be
carried out by using three stencils $S_1$, $S_2$ and $S_3$ that rely
on the variables $\{(B^x)_{J-3/2},(B^x)_{J-1/2},(B^x)_{J+1/2}\}$,
$\{(B^x)_{J-1/2},(B^x)_{J+1/2},(B^x)_{J+3/2}\}$, and
$\{(B^x)_{J+1/2},(B^x)_{J+3/2},(B^x)_{J+5/2}\}$, respectively.  Because the
reconstructed polynomial in the $y$-direction has the form
\begin{align}
&B^x(y)=B^x_0+B^x_y P_1(y) + B^x_{yy} P_2(y), \nonumber
\end{align}
we should calculate $B^x_y$ and $B^x_{yy}$ for the each stencil
in the following manner;
\begin{align}
& (B^x_y)^{(1)}  = \frac{3(B^x)_{J+1/2}-4(B^x)_{J-1/2}+(B^x)_{J-3/2}}{2\Delta y_l}\nonumber\\
& (B^x_{yy})^{(1)} = \frac{(B^x)_{J+1/2}-2(B^x)_{J-1/2}+ (B^x)_{J-3/2}}{2 \Delta y^2_l}
\end{align}
for the stencil $S_1$,
\begin{align}
& (B^x_y)^{(2)}  = \frac{(B^x)_{J+3/2}-(B^x)_{J-1/2}}{2\Delta y_l}\nonumber\\
& (B^x_{yy})^{(2)} = \frac{(B^x)_{J+3/2}-2(B^x)_{J+1/2}+ (B^x)_{J-1/2}}{2 \Delta y^2_l}
\end{align}
for the stencil $S_2$, and
 \begin{align}
& (B^x_y)^{(3)}  = \frac{-(B^x)_{J+5/2}+4(B^x)_{J+3/2}-3(B^x)_{J+1/2}}{2\Delta y_l}\nonumber\\
& (B^x_{yy})^{(3)} = \frac{(B^x)_{J+5/2}-2(B^x)_{J+3/2}+(B^x)_{J+1/2}}{2 \Delta y^2_l}
\end{align}
for the stencil $S_3$, respectively. According to the prescription in
Ref.~\cite{Jiang:1996}, we calculate the weight $\omega^{(k)}$ for
each stencil with $k=1$, 2, and 3. Then, we evaluate the coefficients as
\begin{align}
& B^x_y = \omega^{(1)} (B^x_y)^{(1)} + \omega^{(2)} (B^x_y)^{(2)}
+ \omega^{(3)} (B^x_y)^{(3)} \nonumber \\
& B^x_{yy} = \omega^{(1)} (B^x_{yy})^{(1)} + \omega^{(2)}
(B^x_{yy})^{(2)} + \omega^{(3)} (B^x_{yy})^{(3)}.
\label{eq:weno}
\end{align}
The weight $\omega^{(k)}$ is reduced to be nearly zero if the stencil
$k$ contains a discontinuity, while, for the smooth profile, it is
reduced to be the optimal weight, with which the right-hand side of
Eq.~(\ref{eq:weno}) can be a fifth-order accurate expression of the
derivative. The coefficients $B^x_z$ and $B^x_{zz}$ as well as the
cross term $B^x_{yz}$ are obtained in the similar way.

Essentially the same procedure is applied to the surface at
$x=x_{I+1}$.  To reconstruct $\tilde{B}^y$ ($\tilde{B}^z$),
the permutation rule of $x\to y$, $y\to z$, and $z\to x$ ($x\to z$, $y
\to x$, and $z \to y$) should be simply applied to.

Second, we reconstruct the magnetic field in the interior of the cell,
for which the third-order accurate form is
\begin{align}
\hat{B}^x(x,y,z)&=a_0^x + a_x^x P_1(x) + a_y^x P_1(y) + a_z^x P_1(z)
\nonumber\\
&+ a_{xx}^x P_2(x) + a_{xy}^x P_1(x) P_1(y) + a_{xz}^x
P_1(x) P_1(z) \nonumber\\
& + a_{yy}^x P_2(y)+ a_{yz}^x P_1(y) P_1(z)
+ a_{zz}^x P_2(z) \nonumber\\
&+ a_{xxx}^x P_3(x)+ a_{xxy}^x P_2(x)
P_1(y) + a_{xxz}^x P_2(x) P_1(z) \nonumber\\
&+ a_{xyy}^x P_1(x) P_2(y)
+ a_{xzz}^x P_1(x) P_2(z) \nonumber\\
&+ a_{xyz}^x P_1(x) P_1(y) P_1(z), \label{eq:vol}
\end{align}
where $P_3(x)=(x-x_{I+1/2})^3-3(x-x_{I+1/2})\Delta x_l^2/20$. The
expression for $\hat{B}^y(x,y,z)$ ($\hat{B}^z(x,y,z)$) is also
obtained from the permutation of $x\to y$, $y\to z$, and $z\to x$
($x\to z$, $y \to x$, and $z \to y$). Hence, we have to determine in
total 48 unknown coefficients.  Imposing that $\partial_a \bB^a=0$
holds everywhere inside the cell, we obtain 10 algebraic
equations. Furthermore, we require that the profile for the interior
of the cell matches that on the cell surfaces. Then, 36 algebraic
equations are obtained.
These 46 algebraic equations are not independent, i.e., one of them
can be derived from the others. This implies that there are three
degrees of freedom.  We fix these degrees of freedom in the following
manner: Consider one of the algebraic equations
\begin{align}
 a^x_{xxy} + a^y_{xyy} = -\frac{a^z_{xyz}}{2}, \label{eq:free}
\end{align}
where $a^y_{xyy}~(a^z_{xyz})$ is a coefficient in
$\hat{B}^y(x,y,z)~(\hat{B}^z(x,y,z))$ in the analogy with
Eq.~(\ref{eq:vol}). $a^z_{xyz}$ can be determined by the matching at
the cell surface and in the interior, and there is no equation to
determine $a^x_{xxy}$ and $a^y_{xyy}$ other than Eq.~(\ref{eq:free}).
We follow Ref.~\cite{Balsara:2001rw} to determine these coefficients.
By minimizing the magnetic energy involved in the cell with respect to
$a^x_{xxy}$ and $a^y_{xyy}$, we obtain
\begin{align}
a^x_{xxy} = a^y_{xyy} = - \frac{a^z_{xyz}}{4}; \nonumber
\end{align}
(see Ref.~\cite{Balsara:2001rw} in details). The same procedure is
applied to the coefficients with the permutation of $x\to y$, $y\to z$,
and $z\to x$, and $x\to z$, $y \to x$, and $z \to y$. As a result,
three degrees of freedom are fixed.

Finally, using the algebraic equation~(\ref{eq:vol}) that holds in the
whole interior of the parent cell, the magnetic fields in the eight
cells of the child domain $l+1$, contained within the parent cell $l$,
are reconstructed. Because the algebraic form of the magnetic field
satisfies the divergence-free condition, the magnetic field in the
child cells thus determined satisfies this condition automatically.

\begin{figure*}
 \begin{center}
  \begin{tabular}{c}
    \includegraphics[width=12.0cm,angle=0]{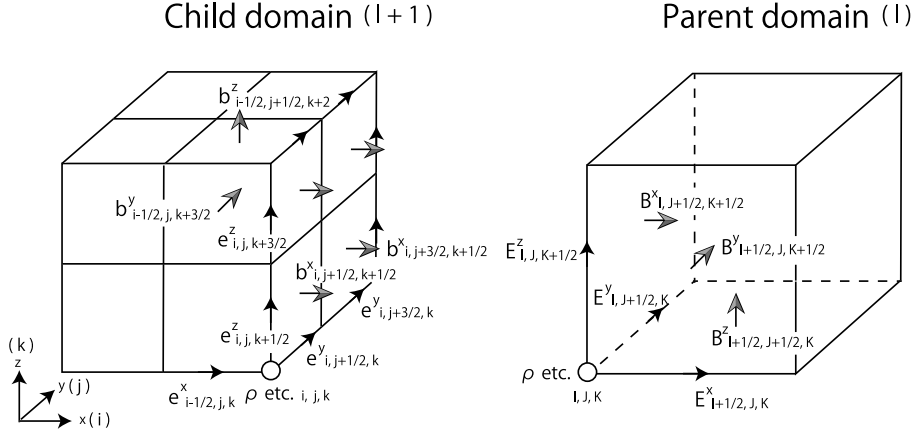}
  \end{tabular}
  \caption{\label{fig:cellB} Schematic picture for the structure of
the cells in our FMR algorithm together with the assigned locations
for the magnetic field in a child domain (left) and in a parent domain
(right). Geometrical and fluid variables are defined at the cell
corner, magnetic field on the cell surface, and flux at the cell edge,
respectively.  }
 \end{center}
 \end{figure*}

\begin{figure*}
 \begin{center}
  \begin{tabular}{c}
    \includegraphics[width=12.0cm,angle=0]{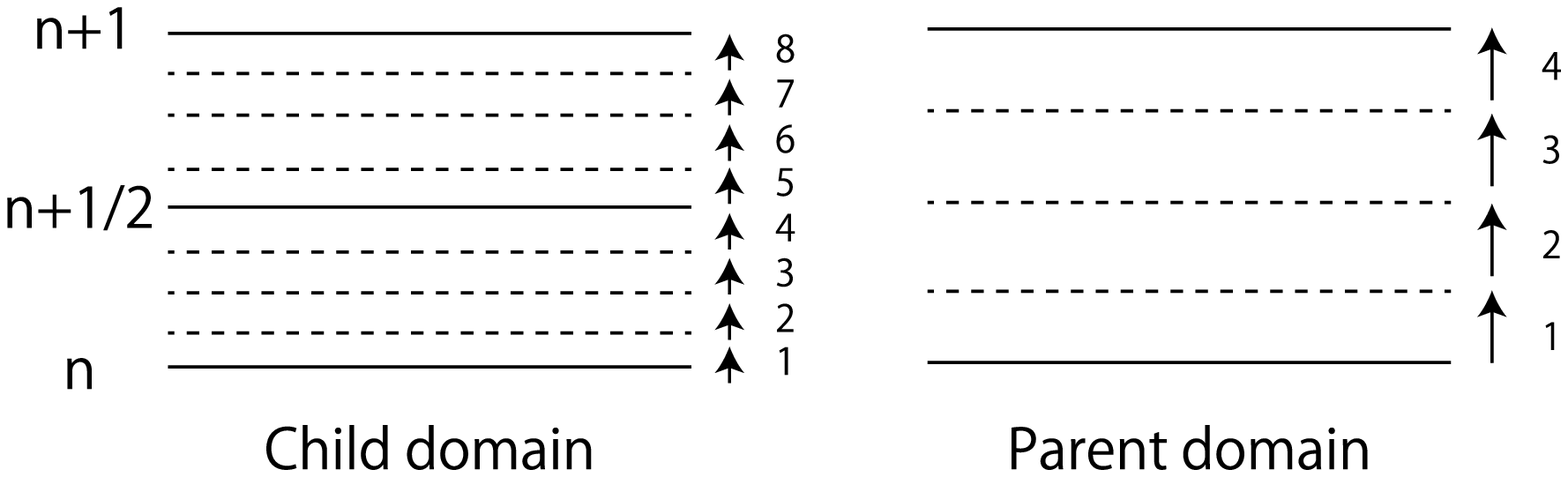}
  \end{tabular}
  \caption{\label{fig:tstep} Schematic picture of the time
  integration scheme for a child domain of label $l+1$ (left) and
  parent domain of label $l$ from $n$ to $n+1$ time slice for the domain
  $l>l_c$.  Attached small arrows with numbers denote the sub-step
  for the fourth-order Runge-Kutta integration.  }
 \end{center}
 \end{figure*}

The restriction of the magnetic field from the child cells to their
parent cell is done at specific time step levels: We choose the
time-step levels in the FMR algorithm following
Refs.~\cite{Berger:1983,Yamamoto:2008js}. Specifically, the time step
intervals for each FMR level is chosen by
\begin{align}
 \Delta t_l =\left\{
 \begin{array}{ll}
 c_{\rm CFL} \Delta x_{l_c}~\text{for }1\le l \le l_c, \\
 c_{\rm CFL} \Delta x_{l}~\text{for }l_c < l \le l_{\rm max},\\
 \end{array}
 \right.
\end{align}
where $c_{\rm CFL}$ is the Courant number $\approx 0.4$ -- 0.5.
Namely, for the coarser levels with $l \leq l_c$, the time step
intervals are chosen to be identical while it is chosen to be
proportional to the grid spacing for $l > l_c$ (see
Fig.~\ref{fig:tstep}). In this setup, the restriction is done when the
time slice of the child domain agrees with that of the corresponding
parent domain (see the time step level $n+1$ in
Fig.~\ref{fig:tstep}). The simplest form for the restriction would be
(cf. Fig.~\ref{fig:cellB})
\begin{align}
(B^x)_{I,J+\frac{1}{2},K+\frac{1}{2}}
=\frac{1}{4} &\Big(b^x_{i,j+\frac{1}{2},k+\frac{1}{2}}
+b^x_{i,j+\frac{3}{2},k+\frac{1}{2}}\nonumber\\
+&b^x_{i,j+\frac{1}{2},k+\frac{3}{2}}
+b^x_{i,j+\frac{3}{2},k+\frac{3}{2}}\Big). \label{eq:restrict}
\end{align}
However, this cannot be employed, because the divergence-free
condition is not satisfied in the parent level: Note that the
divergence-free condition for $B^a$ is preserved if the flux
calculated from the electric field $E^{a}$ is used to integrate the
induction equation. However, in the restriction~(\ref{eq:restrict}),
$e^y$ and $e^z$, instead of $E^y$ and $E^z$, are used to update $B^x$
(cf. Fig.~\ref{fig:cellB}). For such cases, simple restriction schemes
in general do not work well.

Thus, following Ref.~\cite{Balsara:2001rw}, we add a correction in
addition to the ``zeroth-order'' restriction (\ref{eq:restrict}), to
preserve the divergence-free condition of the magnetic field.
Reference~\cite{Balsara:2001rw} proposed to use the following
restriction: For $l > l_c$
\begin{widetext}
\begin{align}
&(B^{y})_{I+\frac{1}{2},J,K+\frac{1}{2}}
\to (B^{y})_{I+\frac{1}{2},J,K+\frac{1}{2}}
+\sum^4_{m=1} \Delta t_l^{(m)}
\frac{(E^z)^{(m)}_{I,J,K+\frac{1}{2}}}{\Delta x_l}
-\frac{1}{4}\sum^8_{m=1}\Delta t_{l+1}^{(m)}
\frac{(e^z)^{(m)}_{i,j,k+\frac{1}{2}}
+(e^z)^{(m)}_{i,j,k+\frac{3}{2}}}{\Delta x_{l+1}}, \nonumber\\
&(B^{y})_{I+\frac{1}{2},J+1,K
+\frac{1}{2}} \to (B^{y})_{I+\frac{1}{2},J+1,K+\frac{1}{2}}
+\sum^4_{m=1} \Delta t_l^{(m)}
\frac{(E^z)^{(m)}_{I,J+1,K+\frac{1}{2}}}{\Delta x_l}
-\frac{1}{4}\sum^8_{m=1} \Delta t_{l+1}^{(m)}
\frac{(e^z)^{(m)}_{i,j+2,k+\frac{1}{2}}
+(e^z)^{(m)}_{i,j+2,k+\frac{3}{2}}}{\Delta x_{l+1}},\nonumber\\
&(B^{z})_{I+\frac{1}{2},J+\frac{1}{2},K}
\to (B^{z})_{I+\frac{1}{2},J+\frac{1}{2},K}
-\sum^4_{m=1} \Delta t_l^{(m)}
\frac{(E^y)^{(m)}_{I,J+\frac{1}{2},K}}{\Delta x_l}
+\frac{1}{4}\sum^8_{m=1} \Delta t_{l+1}^{(m)}
\frac{(e^y)^{(m)}_{i,j+\frac{1}{2},k}
+(e^y)^{(m)}_{i,j+\frac{3}{2},k}}{\Delta x_{l+1}},\nonumber\\
&(B^{z})_{I+\frac{1}{2},J+\frac{1}{2},K+1}
\to (B^{z})_{I+\frac{1}{2},J+\frac{1}{2},K+1}
-\sum^4_{m=1} \Delta t_l^{(m)}
\frac{(E^y)^{(m)}_{I,J+\frac{1}{2},K}}{\Delta x_l}
+\frac{1}{4}\sum^8_{m=1} \Delta t_{l+1}^{(m)}
\frac{(e^y)^{(m)}_{i,j+\frac{1}{2},k+2}
+(e^y)^{(m)}_{i,j+\frac{3}{2},k+2}}{\Delta x_{l+1}},
\label{eq:offset}
\end{align}
\end{widetext}
where the time step intervals of the Runge-Kutta integration are
defined as follows: $\Delta t_{l'}^{(1)}=\Delta t_{l'}^{(4)}=\Delta
t_{l'}/6$, $\Delta t_{l'}^{(2)}=\Delta t_{l'}^{(3)}=\Delta t_{l'}/3$
with $l'=l,l+1$, $\Delta t_{l+1}^{(5)}=\Delta t_{l+1}^{(8)}=\Delta
t_{l+1}/6$, and $\Delta t_{l+1}^{(6)}=\Delta t_{l+1}^{(7)}=\Delta
t_{l+1}/3$.  $(E^y)^{(m)}$, $(E^z)^{(m)}$, $(e^y)^{(m)}$, and
$(e^z)^{(m)}$ denote the electric-field components at sub-step levels,
$m$, of the Runge-Kutta integration (see Fig.~\ref{fig:tstep}). For
$l\le l_c$ for which the time step intervals are identical, the third
term of the right-hand sides of (\ref{eq:offset}) should be replaced
from $\sum^8_{m=1}$ to $\sum^4_{m=1}$. The similar procedure is
applied for $B^y$ and $B^z$ by the permutation of the indices.  These
prescriptions guarantee both the magnetic-flux conservation and the
preservation of the divergence-free condition.

\begin{table*}
\centering
\begin{minipage}{140mm}
\caption{\label{tab:BandE} Grid points where the geometrical and fluid
variables, the magnetic field, and the electric field are defined,
respectively.  }
\begin{tabular}{cc}
\hline\hline
Metric and fluid variables & $(I,J,K)$ \\
$B^x$ & $(I,J+\frac{1}{2},K+\frac{1}{2})$\\
$B^y$ & $(I+\frac{1}{2},J,K+\frac{1}{2})$\\
$B^z$ & $(I+\frac{1}{2},J+\frac{1}{2},K)$\\
$E^x$ & $(I+\frac{1}{2},J,K)$\\
$E^y$ & $(I,J+\frac{1}{2},K)$\\
$E^z$ & $(I,J,K+\frac{1}{2})$\\
\hline\hline
\end{tabular}
\end{minipage}
\end{table*}


\section{Code tests}

\subsection{One-dimensional tests}

We here report the results for one-dimensional MHD tests in the
Minkowski spacetime, proposed in Ref.~\cite{Komissarov:1999}.  The
initial data of the tests are summarized in Table~\ref{tab:test}.  For
all the initial data, a discontinuity is present at $x=0$, and the
left ($x<0$) and right ($x>0)$ states are composed of uniform
profiles.  For all the cases, the $\Gamma$-law EOS with $\Gamma=4/3$
is adopted. We performed these tests in a three-dimensional code
assuming that all the quantities are uniformly distributed in the $y$-
and $z$-directions. The divergence-free condition of the magnetic
field implies that $B^x$ is a constant along the $x$ direction.
Because $B^x$ does not evolve in this test, the divergence-free
condition is automatically satisfied.

\begin{figure*}
 \begin{center}
 \begin{tabular}{cc}
   \includegraphics[width=8.0cm,angle=0]{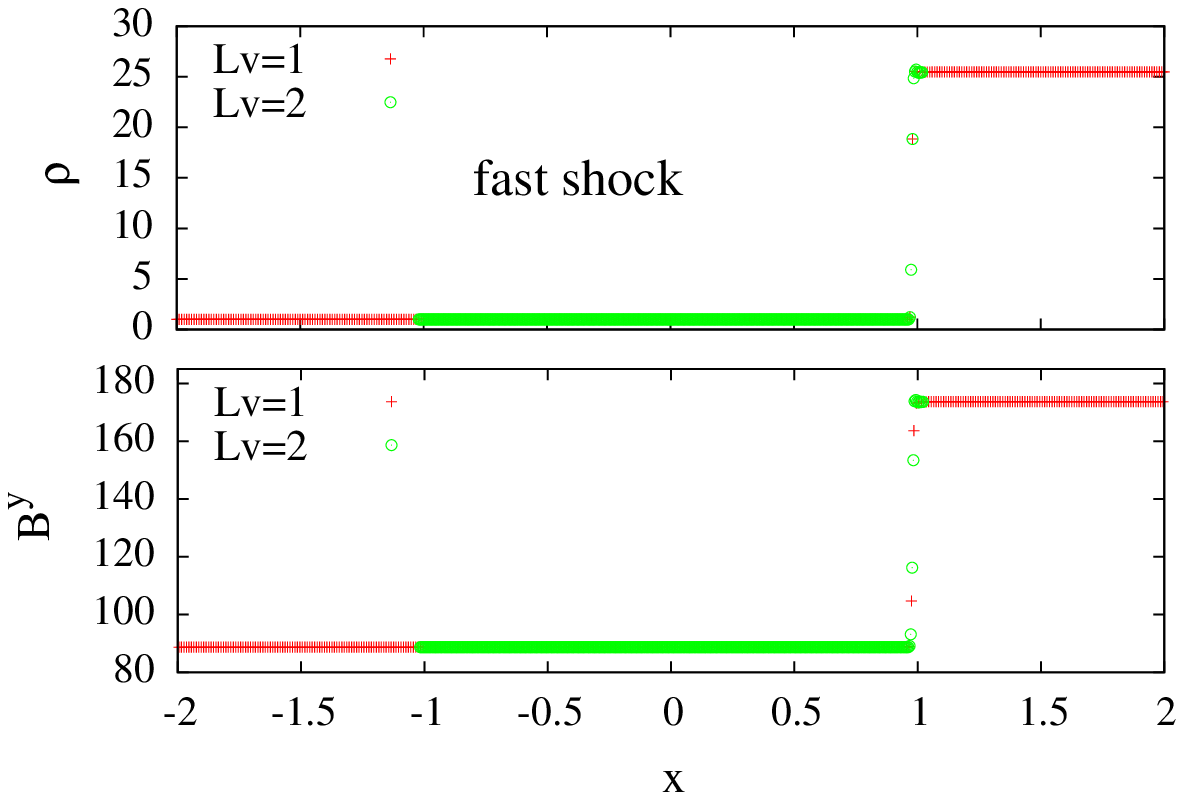} &
   \includegraphics[width=8.0cm,angle=0]{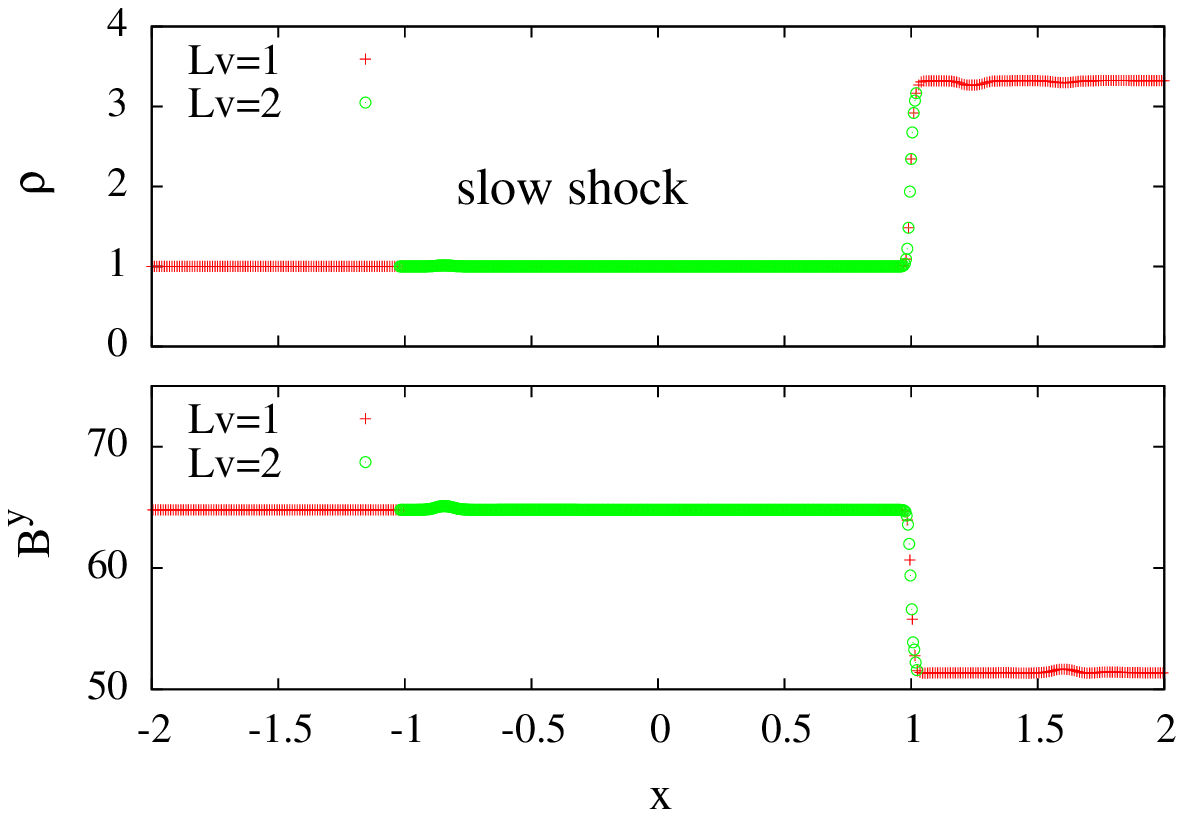}
 \end{tabular}
 \caption{\label{figa1} Snapshots of rest-mass density and
 $y$-component of the magnetic field in 1D fast shock (left) and slow
 shock (right) problems. Numerical solutions are plotted both for the
 FMR domain 1 (red-cross symbol) and domain 2 (green-circle symbol).
 }
 \end{center}
\end{figure*}

\begin{figure*}
 \begin{center}
 \begin{tabular}{cc}
   \includegraphics[width=8.0cm,angle=0]{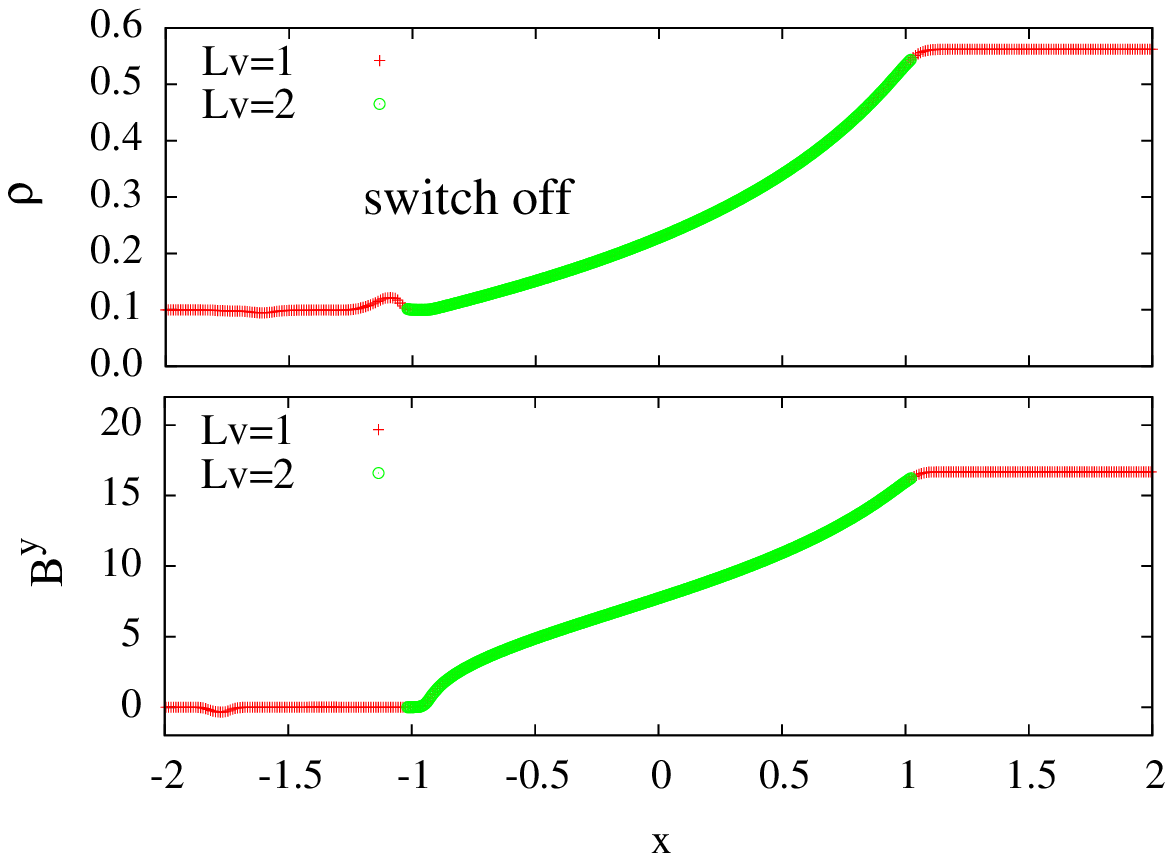} &
   \includegraphics[width=8.0cm,angle=0]{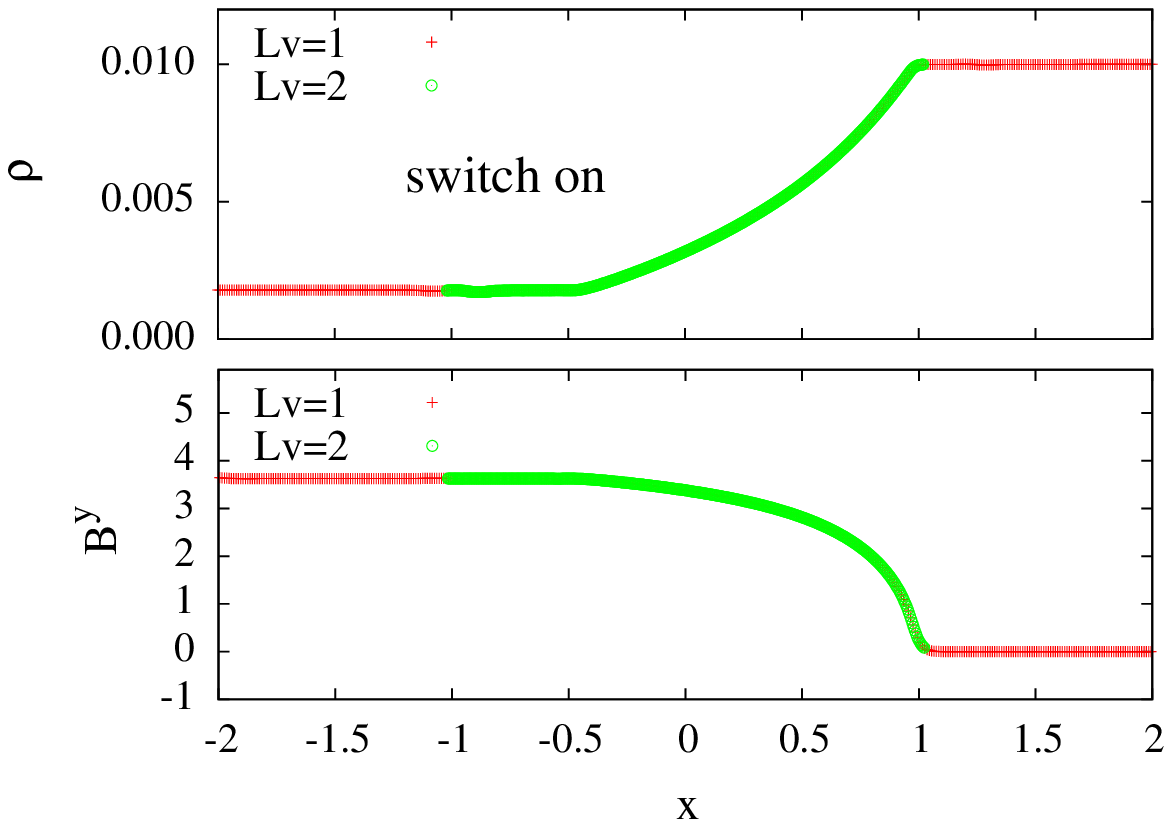}
 \end{tabular}
 \caption{\label{figa2} The same as Fig.~\ref{figa1} but for the 1D
 switch-off and switch-on test problems.  }
 \end{center}
\end{figure*}

\begin{figure*}
 \begin{center}
 \begin{tabular}{cc}
   \includegraphics[width=8.0cm,angle=0]{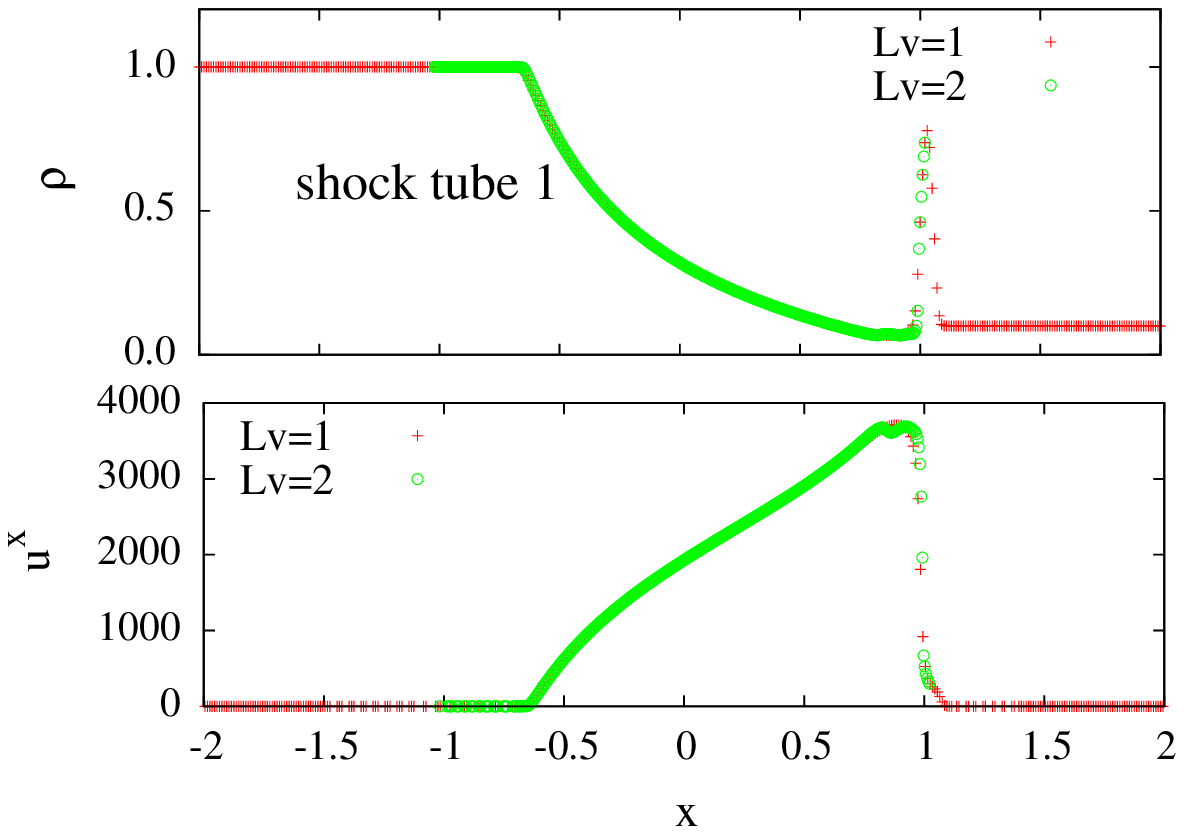} &
   \includegraphics[width=8.0cm,angle=0]{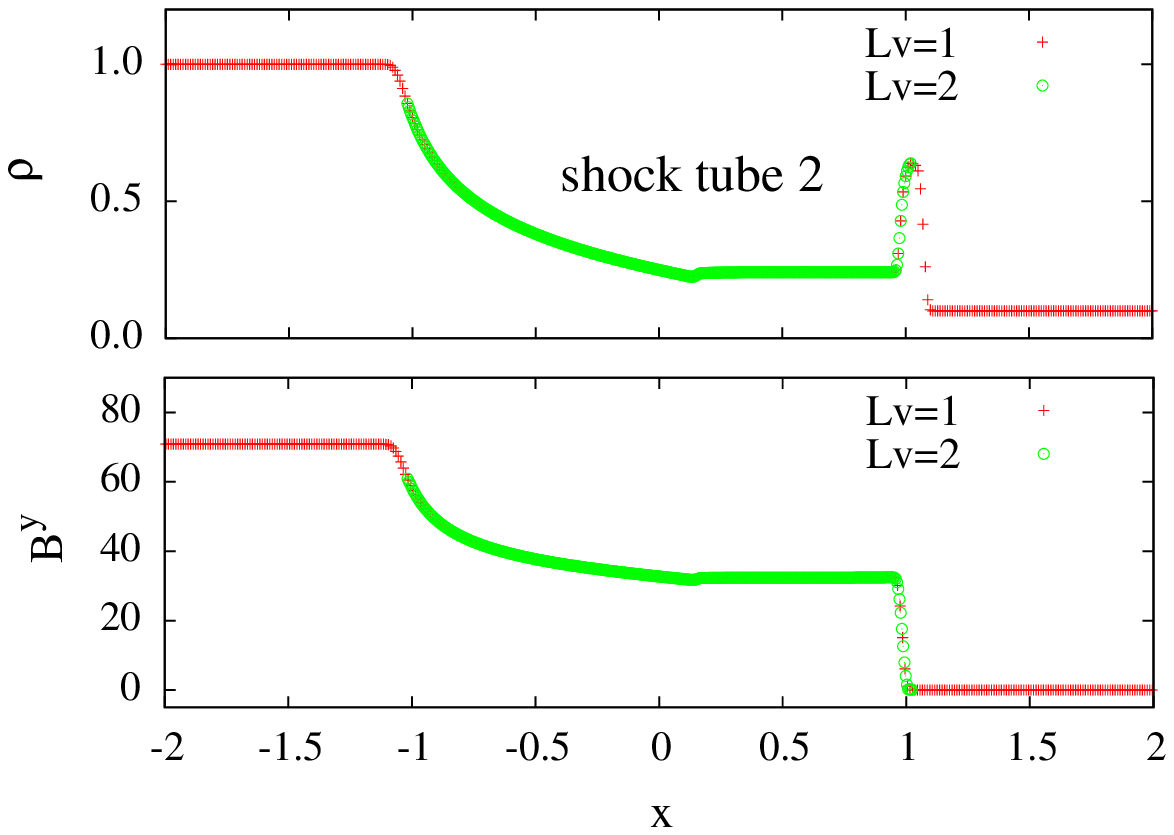}
 \end{tabular}
 \caption{\label{figa3} The same as Fig.~\ref{figa1} but for the
 shock tube 2 tests.  For the 1D shock tube 1, we plot the four
 velocity weighted by the enthalpy instead of the magnetic field
 because $B^y=0$ in this simulation.  }
 \end{center}
\end{figure*}

The purpose of this section is to demonstrate that our FMR code works
well even in the presence of discontinuities and shock waves across
the refinement boundaries. To do this, we prepare a computational
region composed of two FMR domains, for which the grid point and
resolution are summarized in Table~\ref{tab:test}.  The simulations
were terminated at $t_{\rm fin}$ when a discontinuity or waves go
through the refinement boundary.  Figures~\ref{figa1} --
\ref{figa3} plot the snapshots of $\rho$ and $B^y$ at $t=t_{\rm
fin}$. Numerical solutions in the FMR domains 1 and 2 are plotted
together with the red-plus and green-circle symbols. The solutions in
both domains agree approximately with the analytic solutions, except
for a spurious small modulation for $\rho$ in the region $1.0\lesssim
x \lesssim 1.7$ in the slow shock test and a small bump around $x=-1$
in the switch-off test.  As reported in Ref.~\cite{Shibata:2005gp},
these errors are generated by the initially discontinuous data at
$x=0$ irrespective of grid resolutions and presence of the FMR
refinement boundary. The amplitudes of those errors are known to
gradually decrease with improving the grid resolution. The point to be
stressed is that the discontinuities and waves pass thorough from the
domain 2 to domain 1 without any problems.  These results validate the
implementation of staggered magnetic fields, prolongation, and
restriction described in the previous section.

\subsection{Two-dimensional tests}

\begin{figure*}
 \begin{center}
   \includegraphics[width=16.0cm,angle=0]{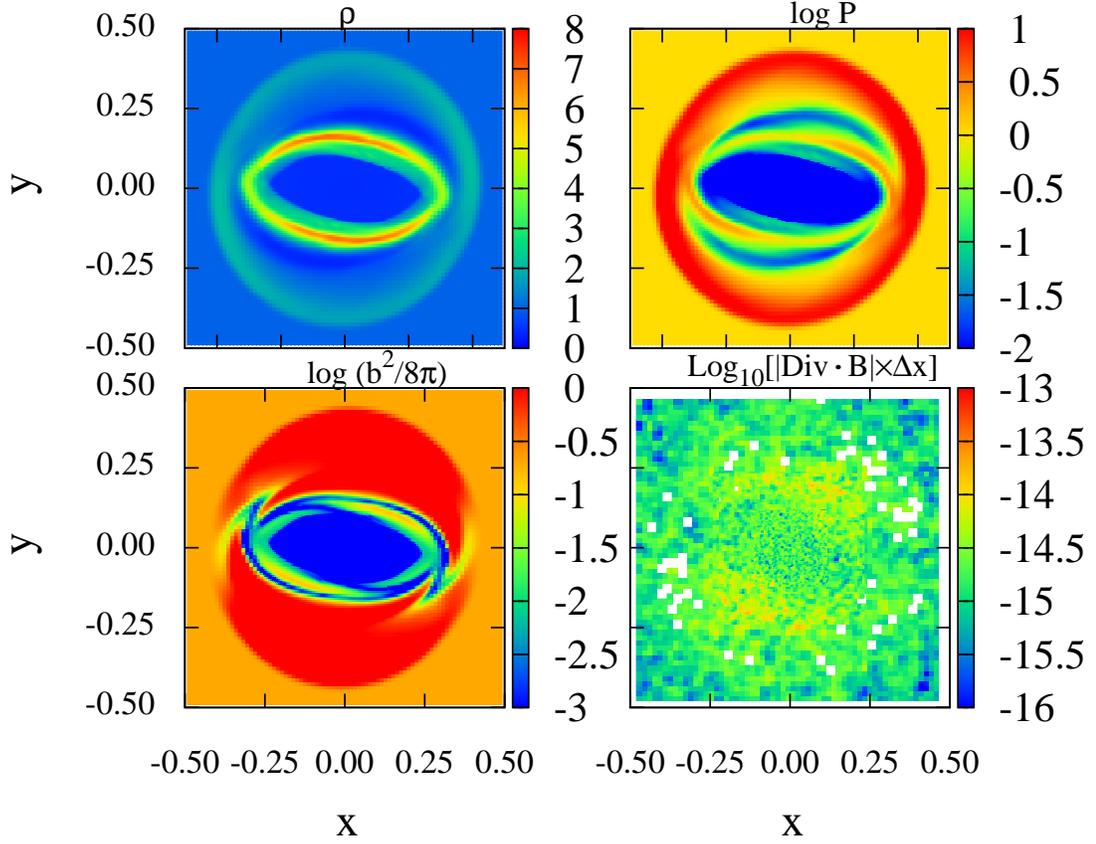}
 \caption{\label{figa4} The results of a cylindrically rotating disk
 problem: Rest-mass density $\rho$ (top-left), pressure (top-right),
 magnetic field strength $b^2$ (bottom-left), and divergence-free
 condition (bottom-right) on the $x$-$y$ plane at $t=0.4$.  }
 \end{center}
\end{figure*}

We also performed a two-dimensional cylindrically rotating disk test
proposed in Ref.~\cite{Delzanna:2003}.  All the variables are assumed
to be functions only of $x$ and $y$ in this test, although the the
simulation was performed by a three-dimensional code.  The employed
initial condition was
\begin{align}
& (\rho,P,B^x/\sqrt{4\pi},B^y/\sqrt{4\pi},v^x,v^y) \nonumber\\
& =\left\{
 \begin{array}{ll}
 (10,1,1,0,-\omega y,\omega x)~{\rm for}~(\sqrt{x^2+y^2}\le 0.1)\\
 (1,1,1,0,0,0)~~~~~~~~~{\rm for}~(\sqrt{x^2+y^2}\ge 0.1)\\
 \end{array}
 \right.
\end{align}
with $\omega=9.95$. In this test, we adopted the $\Gamma=5/3$ EOS and
prepared three FMR domains which are composed of squares with the
regions $x,y \in [-0.512,0.512]$ for domain 1, $x,y \in
[-0.256,0.256]$ for domain 2, and $x,y \in [-0.128,0.128]$ for domain
3.  The finest grid spacing and grid points were chosen to be 0.001
and 128, respectively, which is equivalent to the middle resolution in
Ref.~\cite{Etienne:2010ui}.  Comparing the result in
Ref.~\cite{Etienne:2010ui}, we found that all the quantities were well
reproduced.  The divergence-free condition was also satisfied with a
high precision (see the bottom-right panel of
Fig.~\ref{figa4}). Therefore, we confirm that our FMR implementation
works well.

\begin{table*}
\centering
\begin{minipage}{140mm}
\caption{\label{tab:test} Initial data and grid setup for 1D MHD
tests. Initial states are separated in the left ($x<0$) and right
($x>0$) state with a discontinuity at $x=0$. $2N+1$ and $l_{\rm max}$
denote the number of the grid point covering the interval $[-N\Delta x_l,N\Delta x_l]$
in a FMR domain and the number of FMR domains, respectively. }
\begin{tabular}{cccccccc}
\hline\hline
Test & Left state $(x<0)$ & Right state $(x>0)$ & $t_{\rm fin}$ & $\Delta x_{l_{\rm max}}$ & N & $l_{\rm max}$\\
\hline\hline
Fast Shock & $u^a=(25.0,0.0,0.0)$ & $u^a=(1.091,0.3923,0.00)$ & 4.9 & 0.005 & 204 & 2\\
          & $B^a/\sqrt{4\pi}=(20.0,25.02,0.0)$ & $B^a/\sqrt{4\pi}=(20.0,49.0,0.0)$ & \\
          & $P = 1.0,~\rho=1.0$ & $P=367.5,~\rho=25.48$ & \\
\hline
Slow Shock & $u^a=(1.53,0.0,0.0)$ & $u^a=(0.9571,-0.6822,0.00)$ & 2.0 & 0.005 & 204 & 2\\
          & $B^a/\sqrt{4\pi}=(10.0,18.28,0.0)$ & $B^a/\sqrt{4\pi}=(10.0,14.49,0.0)$ & \\
          & $P = 10.0,~\rho=1.0$ & $P=55.36,~\rho=3.323$ & \\
\hline
Switch-off & $u^a=(-2.0,0.0,0.0)$ & $u^a=(-0.212,-0.590,0.0)$ & 1.8 & 0.005 & 204 & 2\\
Fast Rarefaction & $B^a/\sqrt{4\pi}=(2.0,0.0,0.0)$ & $B^a/\sqrt{4\pi}=(2.0,4.71,0.0)$ & \\
                & $P = 1.0,~\rho=0.1$ & $P=10.0,~\rho=0.562$ & \\
\hline
Switch-on & $u^a=(-0.765,-1.386,0.0)$ & $u^a=(0.0,0.0,0.0)$ & 2.2 & 0.005 & 204 & 2\\
Fast Rarefaction & $B^a/\sqrt{4\pi}=(1.0,1.022,0.0)$ & $B^a/\sqrt{4\pi}=(1.0,0.0,0.0)$ & \\
                & $P = 0.1,~\rho=1.78\times 10^{-3}$ & $P=1.0,~\rho=0.01$ & \\
\hline
Shock Tube 1 & $u^a=(0.0,0.0,0.0)$ & $u^a=(0.0,0.0,0.0)$ & 2.2 & 0.005 & 204 & 2\\
            & $B^a/\sqrt{4\pi}=(1.0,0.0,0.0)$ & $B^a/\sqrt{4\pi}=(1.0,0.0,0.0)$ & \\
            & $P = 1000.0,~\rho=1.0$ & $P=1.0,~\rho=0.1$ & \\
\hline
Shock Tube 2 & $u^a=(0.0,0.0,0.0)$ & $u^a=(0.0,0.0,0.0)$ & 2.3 & 0.005 & 204 & 2\\
            & $B^a/\sqrt{4\pi}=(0.0,0.20,0.0)$ & $B^a/\sqrt{4\pi}=(0.0,0.0,0.0)$ & \\
            & $P = 30.0,~\rho=1.0$ & $P=1.0,~\rho=0.1$ & \\
\hline\hline
\end{tabular}
\end{minipage}
\end{table*}

\end{appendix}



\end{document}